\documentclass[journal]{IEEEtran}
\usepackage{graphicx}
\usepackage{amsmath}
\usepackage{amsfonts}
\usepackage{amssymb}
\usepackage{amsthm}
\usepackage{tikz,pgfplots}
\usepackage{subfigure}
\usetikzlibrary{plotmarks}
\usepackage[sans]{dsfont}
\usepackage{float}
\usepackage{stfloats}
\usepackage{mathabx}
\usepackage{topcapt}
\usepackage{xcolor}

\newcommand{\av}{{\bf a}}

\newcommand{\hv}{{\bf h}}

\newcommand{\sv}{{\bf s}}

\newcommand{\wv}{{\bf w}}

\newcommand{\yv}{{\bf y}}

\newcommand{\zerov}{{\bf 0}}

\newcommand{\Am}{{\bf A}}
\newcommand{\Bm}{{\bf B}}
\newcommand{\Cm}{{\bf C}}
\newcommand{\Dm}{{\bf D}}

\newcommand{\Hm}{{\bf H}}
\newcommand{\Id}{{\bf I}}

\newcommand{\Mm}{{\bf M}}

\newcommand{\Qm}{{\bf Q}}
\newcommand{\Rm}{{\bf R}}

\newcommand{\Um}{{\bf U}}
\newcommand{\Wm}{{\bf W}}

\newcommand{\Xm}{{\bf X}}
\newcommand{\Ym}{{\bf Y}}

\newcommand{\Ac}{{\cal A}}
\newcommand{\Bc}{{\cal B}}
\newcommand{\Cc}{{\cal C}}

\newcommand{\Gc}{{\cal G}}

\newcommand{\Mc}{{\cal M}}
\newcommand{\Nc}{{\cal N}}
\newcommand{\Oc}{{\cal O}}

\newcommand{\Rc}{{\cal R}}

\newcommand{\Hcb}{\pmb{\cal H}}

\newtheoremstyle{mystyle}
{}
{}
{\itshape}
{}
{\bf}
{.}
{.5em}
{}

\theoremstyle{mystyle}
\newtheorem{mytheorem}{Theorem}
\newtheorem{mydef}{Definition}
\newtheorem{myprop}{Proposition}
\begin{document}
\title{A New Family of Low-Complexity STBCs for Four Transmit Antennas}
\author{\IEEEauthorblockN{Amr~Ismail$^*$,~\IEEEmembership{IEEE Member}, Jocelyn~Fiorina$^{**}$,~\IEEEmembership{IEEE Member} and~Hikmet~Sari$^{**}$,~\IEEEmembership{IEEE Fellow}}\\
\IEEEauthorblockA{$^*$ Computer, Electrical and Mathematical Sciences and Engineering (CEMSE) Division, King Abdullah University of Science and Technology (KAUST), Thuwal, Makkah Province, Kingdom of Saudi Arabia\\ $^{**}$ Telecommunications Department, SUPELEC, F-91192 Gif-sur-Yvette, France\\
Email: amrismail.tammam@kaust.edu.sa, $\lbrace$jocelyn.fiorina, and hikmet.sari$\rbrace$@supelec.fr\\}}

\maketitle
\begin{abstract}
Space-Time Block Codes (STBCs) suffer from a prohibitively high decoding complexity unless the low-complexity decodability property is taken into consideration in the STBC design. For this purpose, several families of STBCs that involve a reduced decoding complexity have been proposed, notably the multi-group decodable and the fast decodable (FD) codes. Recently, a new family of codes that combines both of these families namely the fast group decodable (FGD) codes was proposed. In this paper, we propose a new construction scheme for rate-1 FGD codes for $2^a$ transmit antennas. The proposed scheme is then applied to the case of four transmit antennas and we show that the new rate-1 FGD code has the lowest worst-case decoding complexity among existing comparable STBCs. The coding gain of the new rate-1 code is optimized through constellation stretching and proved to be constant irrespective of the underlying QAM constellation prior to normalization. Next, we propose a new rate-2 FD STBC by multiplexing two of our rate-1 codes by the means of a unitary matrix. Also a compromise between rate and complexity is obtained through puncturing our rate-2 FD code giving rise to a new rate-3/2 FD code. The proposed codes are compared to existing codes in the literature and simulation results show that our rate-3/2 code has a lower average decoding complexity while our rate-2 code maintains its lower average decoding complexity in the low SNR region. If a {\it time-out} sphere decoder is employed, our proposed codes outperform existing codes at high SNR region thanks to their lower worst-case decoding complexity. 
\end{abstract}

\section{Introduction}
\IEEEPARstart{T}{he} need for low-complexity decodable STBCs is inevitable in the case of high-rate communications over MIMO systems employing a number of antennas higher than two. This is because despite their low decoding complexity that grows only linearly with the size of the underlying constellation, orthogonal STBCs suffer from a severe rate limitation for more than two antennas \cite{HR_ORTH,CLIFF_ORTH}. On the other hand, the full-rate alternatives to orthogonal STBCs namely the Threaded Algebraic Space-Time (TAST) codes \cite{TAST} and perfect codes \cite{PERFECT_STBC} have generally a prohibitively high decoding complexity.

The decoding complexity may be evaluated by different measures, namely the worst-case decoding complexity measure and the average decoding complexity measure. The worst-case decoding complexity is defined as the minimum number of times an exhaustive search decoder has to compute the Maximum Likelihood (ML) metric to optimally estimate the transmitted codeword \cite{FAST,GLOB10}, or equivalently the number of leaf nodes in the search tree if a sphere decoder is employed, whereas the average decoding complexity measure may be numerically evaluated as the average number of visited nodes by a sphere decoder in order to optimally estimate the transmitted codeword \cite{MIMO_VLSI}.

Arguably, the first proposed low-complexity rate-1 code for the case of four transmit antennas is the quasi-orthogonal (QO)STBC originally proposed by H. Jafarkhani \cite{QOSTBC} and later optimized through constellation rotation to provide full-diversity \cite{FULL_DIVERSITY_QOD,OPT_QORTH_ROT}. The QOSTBC partially relaxes the orthogonality conditions by allowing two complex symbols to be jointly detected. Subsequently, rate-1, full-diversity QOSTBCs were proposed for an arbitrary number of transmit antennas that subsume the original QOSTBC as a special case \cite{GEN_QOSTBC}. In this general framework, the \textit{quasi-orthogonality} stands for decoupling the transmitted symbols into two groups of the same size. 

However, STBCs with lower decoding complexity may be obtained through the concept of multi-group decodability laid by S. Karmakar \textit{et al.} in \cite{MULTI_SYMs,MULT_GR}. Indeed, the multi-group decodability generalizes the quasi-orthogonality by allowing the codeword of symbols to be decoupled into more than two groups not necessarily of the same size. Thanks to this approach, one can obtain rate-1, full-diversity $4$-group decodable STBCs for an arbitrary number of transmit antennas \cite{SAST}.

However, due to the strict rate limitation imposed by the multi-group decodability, another family of STBCs namely fast decodable (FD) STBCs \cite{FAST,2TX_MPLX_ORTH,FD_OPT} has been proposed. These codes are conditionally multi-group decodable thus enabling the use of the conditional detection technique \cite{2TX_MPLX_ORTH} in which the ML detection is carried out in two steps. The first step consists of evaluating the ML estimation of the subset of the transmitted symbols which are separable say $\left(x_1,x_2,\ldots,x_k\right)$, conditioned on a given value of the rest of the symbols $\left(\hat{x}_{k+1},\hat{x}_{k+2},\ldots,\hat{x}_{2K}\right)$ that we may note by $\left(x^{\text{ML}}_1,x^{\text{ML}}_2,\ldots,x^{\text{ML}}_k\vert\hat{x}_{k+1},\hat{x}_{k+2},\ldots,\hat{x}_{2K}\right)$. In the second step, the receiver minimizes the ML metric only over all the possible values of $\left(x_{k+1},x_{k+2},\ldots,x_{2K}\right)$.

Recently, STBCs that are at the time multi-group and fast decodable namely the fast group decodable (FGD) codes were proposed \cite{FGD}. These codes are multi-group decodable such that each group of symbols is fast decodable. The contributions of this paper are summarized in the following:
\begin{itemize}
\item We propose a novel systematic construction of rate-1 FGD STBCs for $2^a$ transmit antennas. The rate-1 FGD code for a number of transmit antennas that is not a power of two is obtained by removing the appropriate number of columns from the rate-1 FGD code corresponding to the nearest greater number of antennas that is a power of two (e.g. the rate-1 FGD STBC for three transmit antennas is obtained by removing a single column from the four transmit antennas rate-1 FGD STBC). 
\item We apply our new construction method to the case of four transmit antennas and show that the resulting new 4$\times$4 rate-1 code can be decoded at half the worst-case decoding complexity of the best known rate-1 STBC.
\item The coding gain of the new 4$\times$4 rate-1 code is optimized through constellation stretching \cite{CONST_STRETCHING} and the nonvanishing determinant (NVD) property \cite{CDA} is proven to be achieved by properly choosing the stretching factor. 
\item We propose a new rate-2 STBC by multiplexing two of the new rate-1 codes by the means of a unitary matrix and numerical optimization. Next we achieve a significant reduction of the worst-decoding complexity at the expense of a rate loss by puncturing the rate-2 code to obtain a new rate-3/2 code. It is worth noting that the proposed rate-3/2 and rate-2 codes can be viewed as block orthogonal STBCs \cite{BOSTBC}.
\end{itemize}
We provide comparisons in terms of bit error rate (BER) performance and average complexity through numerical simulations and show that the reduction in worst-case decoding complexity comes at the expense of a negligible performance loss. We also considered a more practical scenario where a {\it time-out} sphere decoder \cite{TIMEOUT_SD} is used and show that our proposed rate-3/2 and rate-2 codes outperform existing codes at high SNR.

The rest of the paper is organized as follows: The system model is defined and the families of low-complexity STBCs are outlined in Section II. In Section III, we propose our scheme for the rate-1 FGD codes construction for the case of $2^a$ transmit antennas. In Section IV, we address the case of four transmit antennas and the coding gain of the new 4$\times$4 rate-1 STBC is optimized. Next, the rate of the proposed code is increased through multiplexing and numerical optimization. Numerical results are provided in Section V, and we conclude the paper in Section VI. The relevant proofs are provided in the Appendices. 
\subsection*{Notations:}
Hereafter, small letters, bold small letters and bold capital letters will designate scalars, vectors and matrices, respectively. If $\Am$ is a matrix, then $\Am^H$, $\Am^T$ and $\text{det}\left(\Am\right)$ denote the Hermitian, the transpose and the determinant of $\Am$, respectively. We define the $\text{vec}(.)$ as the operator which, when applied to  a $m \times n$ matrix, transforms it  into a $mn\times 1$ vector by simply concatenating vertically the columns of the corresponding matrix. The $\otimes$ operator is the Kronecker product and $\delta_{kj}$ is the Kronecker delta. The $\text{sign}(.)$ operator returns 1 if its scalar input is $\geq 0$ and -1 otherwise. The $\text{round}(.)$ operator rounds its argument to the nearest integer. The $\tilde{(.)}$ operator when applied to a complex vector $\av$ returns $\left[\Re\left(\av^T\right),\Im\left(\av^T\right)\right]^T$ and when applied to complex matrix $\Am$ returns $\left[\Re\left( \Am^T\right),\Im\left(\Am^T\right)\right]^T$. For any two integers $a\ \text{and}\ b,\ a\equiv b \left(\text{mod}\ n\right)$ means that $a-b$ is a multiple of $n$ and $\left(m\right)_k$ means $m$ modulo $k$. $\Mc_n$ is the set of $n\times n$ complex matrices. If $\Ac$ is a finite set, then $\vert \Ac \vert$ denotes its cardinality.  
\section{Preliminaries}
The baseband MIMO channel input-output relationship may be described by: 
\begin{equation}
\underset{T\times N_r}{\Ym} =\underset{T\times N_t}{\Xm} \underset{N_t\times N_r}{\Hm} +\underset{T\times N_r}{\Wm}
\label{MIMO}  
\end{equation} 
where $T$ is the codeword signalling period, $N_r$ is the number of receive antennas, $N_t$ is the number of transmit antennas, $\Ym$ is the received signal matrix, $\Xm$ is the code matrix, $\Hm$ is the channel coefficients matrix with entries $h_{kl} \sim \Cc \Nc(0,1)$, and $\Wm$ is the noise matrix with entries $w_{ij} \sim \Cc \Nc(0,N_{0})$. According to the above model, the $t$'th row of $\Xm$ denotes the symbols transmitted through the $N_t$ transmit antennas during the $t$'th channel use while the $n$'th column denotes the symbols transmitted through the $n$'th transmit antenna during the codeword signalling period $T$.

A STBC matrix that encodes $2K$ real symbols can be expressed as a linear combination of the transmitted symbols as \cite{LDSTBC}:
\begin{equation}
\Xm=\sum^{2K}_{k=1} \Am_kx_k
\label{STBC}
\end{equation}
with $x_k\in\mathbb{R}$ and the $\Am_k, k=1,...,2K$ are $T \times N_t$ complex matrices called dispersion or weight matrices that are required to be linearly independent over $\mathbb{R}$. 
A useful manner to express the system model can be directly obtained by replacing $\Xm$ in \eqref{MIMO} by its expression in \eqref{STBC}: 
\begin{equation}
\Ym=\sum^{2K}_{k=1}\left(\Am_k\Hm\right)x_k+\Wm.
\end{equation} 
Applying the $\text{vec}(.)$ operator to the above we obtain:
\begin{equation}
\text{vec}(\Ym)=\sum^{2K}_{k=1}\left(\Id_{N_r}\otimes\Am_k\right)\text{vec}\left(\Hm\right)x_k+\text{vec}(\Wm).
\label{vec}
\end{equation}
where $\Id_{N_r}$ is the $N_r\times N_r$ identity matrix.\\
If $\yv_i$, $\hv_i$ and $\wv_i$ designate the $i$'th columns of the received signal matrix $\Ym$, the channel matrix $\Hm$ and the noise matrix $\Wm$, respectively, then equation \eqref{vec} can be re-written in matrix form as:
\begin{equation}
\underbrace{\begin{bmatrix}\yv_1\\\vdots\\\yv_{N_r}\end{bmatrix}}_{\yv}=\underbrace{\begin{bmatrix}\Am_{1}\hv_1&\dots&\Am_{2K}\hv_1\\\vdots&\vdots&\vdots\\\Am_{1}\hv_{N_r}&\dots&\Am_{2K}\hv_{N_r}\end{bmatrix}}_{\Hcb}\underbrace{\begin{bmatrix}x_1\\\vdots\\x_{2K}\end{bmatrix}}_{\sv}
+\underbrace{\begin{bmatrix}\wv_1\\\vdots\\\wv_{N_r}\end{bmatrix}}_{\wv}.
\label{complex_model}
\end{equation}
A real system of equations can be obtained by separating the real and imaginary parts of the above to obtain:
\begin{equation}
\tilde{\yv}=\tilde{\Hcb}\sv+\tilde{\wv}
\label{real_model}
\end{equation}
where $\yv,\wv \in \mathbb{R}^{2N_rT\times 1}$, and $\tilde{\Hcb}\in\mathbb{R}^{2N_rT\times 2K}$.
Assuming that $N_rT \geq K$, the QR decomposition of $\tilde{\Hcb}$ yields: 
\begin{equation}
\tilde{\Hcb}=\begin{bmatrix}\Qm_1 & \Qm_2\end{bmatrix} \begin{bmatrix}\Rm \\ \zerov \end{bmatrix}
\end{equation}
where $\Qm_1\in \mathbb{R}^{2N_rT\times 2K}$,$\Qm_2\in\mathbb{R}^{2N_rT \times(2N_rT-2K)}$,  $\Qm_i^T\Qm_j=\delta_{ij}\Id,\ i,j\in\left\lbrace 1,2\right\rbrace$, $\Rm$ is a ${2K \times 2K}$ real upper triangular matrix and $\zerov$ is a 
$(2N_rT-2K)\times 2K$ null matrix.
Accordingly, the ML estimate may be expressed as:
\begin{equation}
\sv^{\text{ML}}=\text{arg}\ \underset{\sv \in \Cc}{\text{min}}\Big\Vert \begin{bmatrix} \Qm_1&\Qm_2 \end{bmatrix}^T\tilde{\yv}-\begin{bmatrix}\Rm^T&\zerov^T\end{bmatrix}^T \sv \Big\Vert^2
\end{equation}
which reduces to:
\begin{equation}
\sv^{\text{ML}}=\text{arg}\ \underset{\sv \in \Cc}{\text{min}}\Vert \yv'-\Rm\sv \Vert^2
\label{R}
\end{equation}
where $\Cc$ is the vector space spanned by information vector $\sv$, and $\yv'=\Qm_1^T\tilde{\yv}$.
In the following, we will outline the known families of low-complexity STBCs and the structure of their corresponding $\Rm$ matrices that enable simplified ML detection.
\subsection{Multi-group decodable codes}
Multi-group decodable STBCs are designed to significantly reduce the worst-case decoding complexity by allowing separate detection of disjoint groups of symbols without any loss of
performance. This is achieved iff the ML metric can be expressed as a sum of terms depending on disjoint groups of symbols.
\begin{mydef}A STBC code that encodes $2K$ real symbols is said to be $g$-group decodable if its weight matrices are such that \cite{MULT_GR,MULTI_SYMs}: 
\begin{equation}
\begin{split}
\Am_k^{H}\Am_l+\Am_l^{H}\Am_k&=\pmb{0},\ \forall \Am_k \in \Gc_i,\ \Am_l \in \Gc_j,\\
1\leq i\neq j\leq g,&\ \vert \Gc_i \vert=n_i,\sum^{g}_{i=1}n_i=2K.
\end{split}
\label{g-group}
\end{equation}
where $\Gc_i$ denotes the set of weight matrices associated to the $i$'th group of symbols.
\end{mydef}
\noindent For instance if a STBC that encodes $2K$ real symbols is $g$-group decodable, its worst-case decoding complexity order can be reduced from $M^K$ to $\sum^{g}_{i=1}\sqrt{M}^{n_i}$ with $M$ being the size of the used square QAM constellation. The worst-case decoding complexity order can be further reduced to $\sum^{g}_{i=1}\sqrt{M}^{n_i-1}$ if the conditional detection via hard slicers is employed. In other words, the $g$-group decodability can be seen to split the original tree with $2K-1$ levels to $g$ smaller trees each with $n_i-1$ levels. In the special case of orthogonal STBCs, the worst-case decoding complexity is $\Oc(1)$ as the PAM slicers need only a fixed number of simple arithmetic operations irrespective of the square QAM constellation size. The corresponding upper triangular matrix $\Rm$ will be a block diagonal matrix: 
\begin{equation*}
\Rm=\begin{bmatrix} \Rm_1\ &\zerov\ &\ldots\ &\zerov\\
                    \zerov\ &\Rm_2\ &\ldots\ &\zerov\\
                    \vdots\ & \vdots\ &\ddots\ &\vdots\\
                    \zerov\ &\zerov\  &\ldots\ &\Rm_g \end{bmatrix}
\end{equation*}
where $\Rm_i$ is a $n_i \times n_i$ upper triangular matrix. 
\subsection{Fast decodable codes}
A STBC is said to be fast decodable \cite{FAST} if it is conditionally multi-group decodable.
\begin{mydef}A STBC that encodes $2K$ real symbols is said to be FD if its weight matrices are such that:
\begin{equation}
\begin{split}
\Am_k^{H}\Am_l+\Am_l^{H}\Am_k=\pmb{0},\ \forall \Am_k \in \Gc_i,\ \Am_l \in \Gc_j,\\ 
1\leq i\neq j\leq g,\ \vert \Gc_i\vert=n_i,\sum^{g}_{i=1}n_i=\kappa<2K.
\end{split}
\label{fast}
\end{equation}
\end{mydef}
\noindent The above definition slightly generalizes the definition of FD codes in \cite{FAST} where the FD code is restricted to be conditionally orthogonal, that is $n_1=n_2\ldots =n_g=1$.
The advantage of FD STBCs is that one can resort to the conditional detection to significantly reduce the worst-case decoding complexity. For instance, if a STBC that encodes $2K$ real symbols is FD, its corresponding worst-case decoding complexity order for square QAM constellations is reduced from $\sqrt{M}^{2K-1}$ to $\sqrt{M}^{2K-\kappa}\times \sum^{g}_{i=1}\sqrt{M}^{n_i-1}$. If the FD code is conditionally orthogonal, the worst-case decoding complexity order is reduced to $\sqrt{M}^{2K-\kappa}$. Fast decodability reduces to the splitting of the last $\kappa$ levels of the real sphere decoder tree into $g$ smaller trees each with $n_i-1,\ i=1,\ldots,g$ levels. In the particular case of $n_i=1,\ \forall i=1,\ldots,g$ (i.e. the FD code is in fact conditionally orthogonal), fast decodability reduces to the removal of the last $\kappa$ levels of the real valued tree. The upper triangular matrix $\Rm$ takes the following form:
\begin{equation*}
\Rm=\begin{bmatrix} \Am\ &\Bm\\
                    \zerov\ &\Cm \end{bmatrix}
\end{equation*}
where $\Bm$ has no special structure, $\Cm$ is a $\left(2K-\kappa\right)\times\left(2K-\kappa\right)$ upper triangular matrix, and $\Am$ is a block diagonal $\kappa \times \kappa$ matrix:
\begin{equation*}
\Am=\begin{bmatrix} \Rm_1\ &\zerov\ &\ldots\ &\zerov\\
                    \zerov\ &\Rm_2\ &\ldots\ &\zerov\\
                    \vdots\ & \vdots\ &\ddots\ &\vdots\\
                    \zerov\ &\zerov\  &\ldots\ &\Rm_g \end{bmatrix}
\end{equation*}
with $\Rm_i$ being a $n_i \times n_i$ upper triangular matrix.
\subsection{Fast group decodable codes}
A STBC is said to be fast group decodable \cite{FGD} if it is multi-group decodable such that each group is fast decodable.
\begin{mydef} A STBC that encodes $2K$ real symbols is said to be FGD if:
\begin{equation}
\begin{split}
\Am_k^{H}\Am_l+\Am_l^{H}\Am_k&=\pmb{0},\ \forall \Am_k \in \Gc_i,\ \Am_l \in \Gc_j,\\ 
1\leq i\neq j\leq g,&\ \vert\Gc_i\vert=n_i,\sum^{g}_{i=1}n_i=2K
\end{split}
\end{equation}
and that the weight matrices within each group are such that:
\begin{equation}
\begin{split}
\Am_k^{H}\Am_l+\Am_l^{H}\Am_k&=\pmb{0},\ \forall \Am_k \in \Gc_{i,m},\ \Am_l \in \Gc_{i,n},\\ 
1\leq m\neq n\leq g_i,\ &\vert\Gc_{i,j}\vert=n_{i,j},\ \sum^{g_i}_{j=1}n_{i,j}=\kappa_i<n_i.
\end{split}
\end{equation}
\end{mydef}
\noindent where $\Gc_{i,m}$ (resp. $g_i$) denotes the set of weight matrices that constitute the $m$'th group  (resp. the number of inner groups) within the $i$'th group of symbols $\Gc_i$. For instance, if a STBC that encodes $2K$ real symbols is FGD, its corresponding worst-case decoding complexity order for square QAM constellations is reduced from $\sqrt{M}^{2K-1}$ to $\sum^{g}_{i=1} \sqrt{M}^{n_i-\kappa_i} \times \sum^{g_i}_{j=1}\sqrt{M}^{n_{i,j}-1}$. Similarly, if each group is conditionally orthogonal, the worst-case decoding complexity order is equal to $\sum^{g}_{i=1} \sqrt{M}^{n_i-\kappa_i}$. If a real sphere decoder is employed, fast group decodability reduces to splitting the original $2K-1$ real valued tree into $g$ smaller trees each with $n_i$ levels, and that for each of these new trees the last $\kappa_i$ levels are split into $g_i$ smaller trees each with $n_{i,j}-1$ levels. The corresponding upper triangular matrix $\Rm$ takes the following form:
\begin{equation}
\Rm=\begin{bmatrix} \Rm_1\ &\zerov\ &\ldots\ &\zerov\\
                    \zerov\ &\Rm_2\ &\ldots\ &\zerov\\
                    \vdots\ & \vdots\ &\ddots\ &\vdots\\
                    \zerov\ &\zerov\  &\ldots\ &\Rm_g \end{bmatrix}
\end{equation}
such that for $1\leq i\leq g$, one has: 
\begin{equation}
\Rm_i=\begin{bmatrix} \Am_i\ &\Bm_i\\
                    \zerov\ &\Cm_i \end{bmatrix}
\end{equation}
where $\Bm_i$ has no special structure, $\Cm_i$ is a $\left(n_i-\kappa_i\right)\times\left(n_i-\kappa_i\right)$ upper triangular matrix, and $\Am_i$ is a $\kappa_i \times \kappa_i$ block diagonal matrix:
\begin{equation}
\Am_i=\begin{bmatrix} \Rm_{i,1}\ &\zerov\ &\ldots\ &\zerov\\
                    \zerov\ &\Rm_{i,2}\ &\ldots\ &\zerov\\
                    \vdots\ & \vdots\ &\ddots\ &\vdots\\
                    \zerov\ &\zerov\  &\ldots\ &\Rm_{i,g_i} \end{bmatrix}
\end{equation}
where $\Rm_{i,j}$ being a $n_{i,j} \times n_{i,j}$ upper triangular matrix
\section{The proposed FGD scheme}
In this section, we provide a systematic approach for constructing rate-1 FGD STBCs for $2^a$ transmit antennas. Recognizing the key role of the matrix representations of the generators of the Clifford Algebra over $\mathbb{R}$ \cite{CLIFF_ORTH} in the sequel of this paper, we briefly review their main properties in the following. 
\subsection{Linear Representations of Clifford generators}
The Clifford algebra over $\mathbb{R}$ is generated by the generators $\gamma_i$ satisfying the following property:
\begin{equation}
\gamma_i\gamma_j+\gamma_j\gamma_i=-2\delta_{ij}\mathds{1}
\label{cliff_gen}
\end{equation} 
The above equation can be split in two equations:
\begin{eqnarray}
\gamma_i^2&=&-\mathds{1}\\
\gamma_i\gamma_j&=&-\gamma_j\gamma_i\ \forall i\neq j.
\end{eqnarray}
In \cite{CLIFF_ORTH}, the question about the maximum number of unitary representations of the Clifford generators has been thoroughly addressed and it has been proven that for $2^a \times 2^a$ matrices there are exactly $2a+1$ unitary matrix representations of the Clifford algebra generators. The matrix representations of $\gamma_i$ denoted $\Rc\left(\gamma_i\right)$ for the $2^a \times 2^a$ case are obtained as \cite{CLIFF_ORTH}:
\begin{eqnarray}
\Rc(\gamma_1)&=&\pm j \underbrace{\sigma_3 \otimes \sigma_3\ldots \otimes\sigma_3}_{a}\nonumber\\
\Rc(\gamma_2)&=&\Id_{2^{a-1}}\otimes \sigma_1\nonumber\\
\Rc(\gamma_3)&=&\Id_{2^{a-1}}\otimes \sigma_2\nonumber\\
&\vdots&\nonumber\\
\Rc(\gamma_{2k})&=&\Id_{2^{a-k}}\otimes \sigma_1\underbrace{\otimes \sigma_3\otimes \sigma_3\ldots\otimes \sigma_3}_{k-1}\nonumber\\
\Rc(\gamma_{2k+1})&=&\Id_{2^{a-k}}\otimes \sigma_2\underbrace{\otimes \sigma_3\otimes \sigma_3\ldots\otimes \sigma_3}_{k-1}\nonumber\\
&\vdots&\nonumber\\
\Rc(\gamma_{2a})&=&\sigma_1\otimes \underbrace{\sigma_3 \otimes \sigma_3\ldots\sigma_3}_{a-1}\nonumber\\
\Rc(\gamma_{2a+1})&=&\sigma_2\otimes \underbrace{\sigma_3 \otimes \sigma_3\ldots\sigma_3}_{a-1}
\label{reps}
\end{eqnarray}
where
\begin{equation}
\sigma_1=\begin{bmatrix}0&1\\-1&0\end{bmatrix},\ \sigma_2=\begin{bmatrix}0&j\\j&0\end{bmatrix},\ \sigma_3=\begin{bmatrix}1&0\\0&-1\end{bmatrix}
\label{sigmas}
\end{equation}
From now on, we will denote $\Rc(\gamma_i)$ by $\Rm_i$ for simplicity. The properties of the matrices $\Rm_i,\ i=1,\ldots,2a+1$ can be summarized as follows:
\begin{equation}
\begin{split}
\Rm_i^{H}=-\Rm_i,\ \Rm_i^2=-\Id,\ \forall 1\leq i\leq 2a+1,\\ 
\text{and}\ \Rm_i\Rm_j+\Rm_j\Rm_i=\pmb{0},\ \forall 1\leq i\neq j\leq 2a+1.
\end{split}
\label{properties}
\end{equation}
\begin{table*}[t!]
\centering
\normalsize
\topcaption{Different cases for $\Ac$}
\begin{tabular}{|c|c|}
\hline
$a$&$\Ac$\\
\hline
$4n$&$\left\lbrace j\prod^{2a+1}_{i=a+1}\Rm_i\right\rbrace \overset{a-2}{\underset{m=1}{\cup}} \left\lbrace j^{\delta_{\Ac}(m)}\prod^{2a+1}_{i=a+1}\Rm_i\prod^{m}_{i=1}\Rm_{k_i}:1\leq k_1<\ldots<k_m\leq a \right\rbrace$\\
\hline
$4n+1$&$\left\lbrace \prod^{2a+1}_{i=a+1}\Rm_i\right\rbrace \overset{a-2}{\underset{m=1}{\cup}} \left\lbrace j^{\delta_{\Ac}(m)}\prod^{2a+1}_{i=a+1}\Rm_i\prod^{m}_{i=1}\Rm_{k_i}:1\leq k_1<\ldots<k_m\leq a \right\rbrace$\\
\hline
$4n+2$&$\left\lbrace \prod^{2a+1}_{i=a+1}\Rm_i\right\rbrace \overset{a-2}{\underset{m=1}{\cup}}\left\lbrace  j^{\delta_{\Ac}(m)}\prod^{2a+1}_{i=a+1}\Rm_i\prod^{m}_{i=1}\Rm_{k_i}:1\leq k_1<\ldots<k_m\leq a \right\rbrace$\\
\hline
$4n+3$&$\left\lbrace j\prod^{2a+1}_{i=a+1}\Rm_i\right\rbrace \overset{a-2}{\underset{m=1}{\cup}}\left\lbrace  j^{\delta_{\Ac}(m)}\prod^{2a+1}_{i=a+1}\Rm_i\prod^{m}_{i=1}\Rm_{k_i}:1\leq k_1<\ldots<k_m\leq a \right\rbrace$\\
\hline
\end{tabular}
\label{casesA}
\end{table*}
\begin{table*}[t!]
\centering
\normalsize
\topcaption{Different cases for $\Bc$}
\begin{tabular}{|c|c|}
\hline
$a$&$\Bc$\\
\hline
$4n$&$\left\lbrace j\prod^{a}_{i=1}\Rm_i\right\rbrace \overset{a-2}{\underset{m=2,4}{\cup}}\left\lbrace j^{\delta_{\Bc}(m)}\prod^{a}_{i=1}\Rm_i\prod^{m}_{i=1}\Rm_{k_i}:a+1\leq k_1<\ldots<k_m\leq 2a+1 \right\rbrace$\\
\hline
$4n+1$&$\overset{a-2}{\underset{m=1,3}{\cup}}\left\lbrace j^{\delta_{\Bc}(m)}\prod^{a}_{i=1}\Rm_i\prod^{m}_{i=1}\Rm_{k_i}:a+1\leq k_1<\ldots<k_m\leq 2a+1 \right\rbrace$\\
\hline
$4n+2$&$\left\lbrace \prod^{a}_{i=1}\Rm_i\right\rbrace \overset{a-2}{\underset{m=2,4}{\cup}}\left\lbrace j^{\delta_{\Bc}(m)}\prod^{a}_{i=1}\Rm_i\prod^{m}_{i=1}\Rm_{k_i}:a+1\leq k_1<\ldots<k_m\leq 2a+1 \right\rbrace$\\
\hline
$4n+3$&$\overset{a-2}{\underset{m=1,3}{\cup}}\left\lbrace j^{\delta_{\Bc}(m)}\prod^{a}_{i=1}\Rm_i\prod^{m}_{i=1}\Rm_{k_i}:a+1\leq k_1<\ldots<k_m\leq 2a+1 \right\rbrace$\\
\hline
\end{tabular}
\label{casesB}
\vspace*{0.3cm}
\hrule
\end{table*}
\subsection{A systematic approach for the construction of FGD codes}
\begin{myprop}
For $2^a$ transmit antennas, the two sets of matrices, namely $\Gc_1=\left\lbrace \Id,\Rm_1,\ldots,\Rm_a\right\rbrace\cup\Ac $ and $\Gc_2=\left\lbrace \Rm_{a+1},\ldots,\Rm_{2a+1}\right\rbrace \cup\Bc$ satisfy \eqref{g-group} where $\Ac$ and $\Bc$ are given in Table~\ref{casesA} and Table~\ref{casesB}, respectively, and $\delta_{\Ac}(m), \delta_{\Bc}(m)$ are given in Table~\ref{delta}: 
\end{myprop}
\begin{table}[h!]
\centering
\normalsize
\topcaption{Different cases for $\delta$}
\begin{tabular}{|c|c|c|}
\hline
$a$&$\delta_{\Ac}(m)$&$\delta_{\Bc}(m)$\\
\hline
$4n$&$\frac{\left(\left(m\right)_4-1\right)\left(\left(m\right)_4-2\right)}{2}$&$\frac{2-\left(m\right)_4}{2}$\\
\hline
$4n+1$&$\frac{\left(\left(m\right)_4\left(\left(m\right)_4-1\right)\right)_4}{2}$&$\frac{\left(m\right)_4-1}{2}$\\
\hline
$4n+2$&$\frac{\left(\left(m\right)_4\left(\left(m\right)_4-3\right)\right)_4}{2}$&$\frac{\left(m\right)_4}{2}$\\
\hline
$4n+3$&$\frac{\left(\left(\left(m\right)_4-2\right)\left(\left(m\right)_4-3\right)\right)_4}{2}$&$\frac{3-\left(m\right)_4}{2}$\\
\hline
\end{tabular}
\label{delta}
\end{table}

See Appendix A for proof.
\begin{myprop}
The rate of the proposed FGD codes is equal to one complex symbol per channel use regardless of the number of transmit antennas.
\end{myprop}
See Appendix B for proof.\\
Examples of the rate-1 FGD codes for 4,8 and 16 transmit antennas are given in Table~\ref{examples}.

\begin{table}[h!]
\centering
\normalsize
\topcaption{Examples of rate-1 FGD codes}
\begin{tabular}{|c|c|c|}
\hline
Tx&$\Gc_1$&$\Gc_2$\\
\hline
4&$\Id,\Rm_2,\Rm_4,\Rm_1\Rm_3\Rm_5$&$\Rm_1,\Rm_3,\Rm_5,\Rm_2\Rm_4$\\
\hline
&$\Id,\Rm_2,\Rm_4,\Rm_6$&$\Rm_1,\Rm_3,\Rm_5,\Rm_7$\\
&$j\Rm_1\Rm_3\Rm_5\Rm_7$&$j\Rm_2\Rm_4\Rm_6\Rm_1$\\
8&$j\Rm_1\Rm_3\Rm_5\Rm_7\Rm_2$&$j\Rm_2\Rm_4\Rm_6\Rm_3$\\
&$j\Rm_1\Rm_3\Rm_5\Rm_7\Rm_4$&$j\Rm_2\Rm_4\Rm_6\Rm_5$\\
&$j\Rm_1\Rm_3\Rm_5\Rm_7\Rm_6$&$j\Rm_2\Rm_4\Rm_6\Rm_7$\\
\hline
&$\Id,\Rm_2,\Rm_4,\Rm_6,\Rm_8$&$\Rm_1,\Rm_3,\Rm_5,\Rm_7,\Rm_9$\\
&$j\Rm_1\Rm_3\Rm_5\Rm_7\Rm_9$&$j\Rm_2\Rm_4\Rm_6\Rm_8$\\
&$\Rm_1\Rm_3\Rm_5\Rm_7\Rm_9\Rm_2$&$\Rm_2\Rm_4\Rm_6\Rm_8\Rm_1\Rm_3$\\
&$\Rm_1\Rm_3\Rm_5\Rm_7\Rm_9\Rm_4$&$\Rm_2\Rm_4\Rm_6\Rm_8\Rm_1\Rm_5$\\
&$\Rm_1\Rm_3\Rm_5\Rm_7\Rm_9\Rm_6$&$\Rm_2\Rm_4\Rm_6\Rm_8\Rm_1\Rm_7$\\
16&$\Rm_1\Rm_3\Rm_5\Rm_7\Rm_9\Rm_8$&$\Rm_2\Rm_4\Rm_6\Rm_8\Rm_1\Rm_9$\\
&$\Rm_1\Rm_3\Rm_5\Rm_7\Rm_9\Rm_2\Rm_4$&$\Rm_2\Rm_4\Rm_6\Rm_8\Rm_3\Rm_5$\\
&$\Rm_1\Rm_3\Rm_5\Rm_7\Rm_9\Rm_2\Rm_6,$&$\Rm_2\Rm_4\Rm_6\Rm_8\Rm_3\Rm_7$\\
&$\Rm_1\Rm_3\Rm_5\Rm_7\Rm_9\Rm_2\Rm_8$&$\Rm_2\Rm_4\Rm_6\Rm_8\Rm_3\Rm_9$\\
&$\Rm_1\Rm_3\Rm_5\Rm_7\Rm_9\Rm_4\Rm_6,$&$\Rm_2\Rm_4\Rm_6\Rm_8\Rm_5\Rm_7$\\
&$\Rm_1\Rm_3\Rm_5\Rm_7\Rm_9\Rm_4\Rm_8$&$\Rm_2\Rm_4\Rm_6\Rm_8\Rm_5\Rm_9$\\
&$\Rm_1\Rm_3\Rm_5\Rm_7\Rm_9\Rm_6\Rm_8$&$\Rm_2\Rm_4\Rm_6\Rm_8\Rm_7\Rm_9$\\
\hline
\end{tabular}
\label{examples}
\end{table}
\section{The four transmit antennas case}
In this section, a special attention is given to the case of four transmit antennas. The proposed rate-1 FGD code arises as the direct application of {\bf Proposition 1.} in the case of four transmit antennas. Setting $a=2$, we have $\Gc_1=\left\lbrace \Id,\Rm_2,\Rm_4,\Rm_1\Rm_3\Rm_5\right\rbrace $ and $\Gc_2=\left\lbrace\Rm_1,\Rm_3,\Rm_5,\Rm_2\Rm_4\right\rbrace$. It is worth noting that the above choice of weight matrices guarantees an equal average transmitted power per transmit antenna per time slot.
\subsection{A new rate-1 FGD STBC for four transmit antennas}
The Proposed rate-1 STBC denoted $\Xm_1$ is expressed as:
\begin{equation}
\begin{split}
\Xm_1(\sv)=&\Id x_1+\Rm_2 x_2+\Rm_4 x_3+\Rm_1\Rm_3\Rm_5 x_4+\\
&\Rm_1 x_5+\Rm_3 x_6+\Rm_5 x_7+\Rm_2\Rm_4 x_8.
\end{split}
\end{equation}
According to {\bf Definition 3.}, the proposed code $\Xm_1$ is FGD with $g=2,n_1=n_2=4$ and $g_1=g_2=3$ such as $n_{i,j}=1,\ i=1,2,\ j=1,2,3$. Therefore, the worst-case decoding complexity order is $2\sqrt{M}$. However, the coding gain of $\Xm_1$ is equal to zero, and in order to achieve full-diversity, we resort to the constellation stretching \cite{CONST_STRETCHING} rather than the constellation rotation technique, otherwise the symbols inside each group will be entangled together which in turns will destroy the FGD structure of the proposed code and causes a significant increase in the decoding complexity. 

\begin{figure}[h!]
\centering 
\begin{tikzpicture}[x=0.65cm,y=0.65cm]
\draw[-] (-4,0) -- (4,0) coordinate (x axis);
\draw[-] (0,-4) -- (0,4) coordinate (y axis);
\draw[step=0.75cm,gray!50,very thin] (-4,-4) grid (4,4);
\draw[-stealth,thick] (-5,0) -- (5,0) node[right] {$I$} coordinate (x axis);
\draw[-stealth,thick] (0,-5) -- (0,5) node[right] {$Q$}coordinate (y axis);
\foreach \x in {-3,-1,1,3}
 \foreach \y in {-3,-1,1,3}
   {
     \draw [black] plot [only marks, mark size=1.5, mark=*]coordinates {(\x,\y)};
   }
\foreach \x in {-3.3541,-1.1180,1.1180,3.3541}
 \foreach \y in {-2.5981,-0.8660,0.8660,2.5981}
   {
     \draw [red] plot [only marks, mark size=1.5, mark=square*] coordinates {(\x,\y)};
   }   
\end{tikzpicture}
\caption{regular ($\bullet$) versus stretched ($\color{red}\sqbullet$) 16-QAM constellation points}
\label{stretchedconst}
\end{figure}
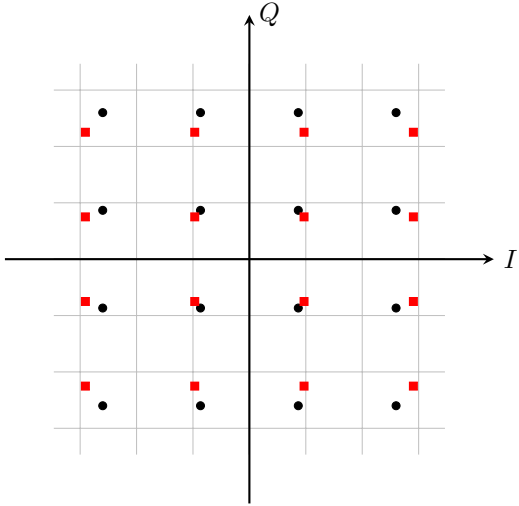

The full diversity code matrix takes the form of \eqref{newcode} where $\sv=[x_1,\ldots,x_8]$ and $k$ is chosen to provide a high coding gain. The term $\sqrt{\frac{2}{1+k^2}}$ is added to normalize the average transmitted power per antenna per time slot.
\begin{table*}[t!]
\normalsize
\begin{equation}
\Xm_1(\sv)=\sqrt{\frac{2}{1+k^2}}\begin{bmatrix}
x_1+ikx_5&x_2+ikx_6&x_3+ikx_7&-ikx_4-x_8\\
-x_2+ikx_6&x_1-ikx_5&-ikx_4-x_8&-x_3-ikx_7\\
-x_3+ikx_7&ikx_4+x_8&x_1-ikx_5&x_2+ikx_6\\
ikx_4+x_8&x_3-ikx_7&-x_2+ikx_6&x_1+ikx_5 
\end{bmatrix} 
\label{newcode} 
\end{equation}
\hrule
\end{table*}
\begin{myprop}
Taking $k=\sqrt{\frac{3}{5}}$, ensures the NVD property for the proposed code with a coding gain equal to 1.
\end{myprop}
See Appendix C for proof.\\
For illustration purposes, a comparison between regular and stretched 16-QAM constellation points is depicted in Fig~\ref{stretchedconst} where the dark dots denote the regular 16-QAM constellation points whereas the red squares denote the stretched 16-QAM constellation points with stretching factor $k=\sqrt{\frac{3}{5}}$.

\subsection{The proposed rate-2 code}
The proposed rate-2 code denoted $\Xm_2$ is simply obtained by multiplexing two rate-1 codes by means of a unitary matrix. Mathematically speaking, the rate-2 STBC is expressed as:
\begin{equation}
\Xm_2\left(x_1,\ldots,x_{16}\right)=\Xm_1\left(x_1,\ldots,x_8\right)+e^{j\phi}\Xm_1\left(x_9,\ldots,x_{16} \right)\Um  
\label{R2}
\end{equation}
where $\Um$ and $\phi$ are chosen in order to maximize the coding gain. It was numerically verified for QPSK constellation that taking $\Um_{\text{opt}}=j\Rm_1$ and $\phi_{\text{opt}}=\tan^{-1}\left(\frac{1}{2}\right)$ maximizes the coding gain which is equal to 1.  

To decode the proposed code, the receiver evaluates the QR decomposition of the real equivalent channel matrix $\tilde{\Hcb}$ \eqref{real_model}. Thanks to the FD structure of the proposed rate-2 code with $K=8,\kappa=8,n_1=n_2=4$, the corresponding upper-triangular matrix $\Rm$ takes the form:
\begin{equation}
\Rm=\begin{bmatrix} \Am& \Bm\\ \zerov& \Cm \end{bmatrix}
\end{equation}
where $\Bm \in \mathbb{R}^{8 \times 8}$ has no special structure, $\Cm \in \mathbb{R}^{8 \times 8}$ is an upper triangular matrix and $\Am \in \mathbb{R}^{8 \times 8}$ takes the form:
\begin{equation}
\Am=\begin{bmatrix} x\ &\ 0\ &\ 0\ &\ x\ &\ 0\ &\ 0\ &\ 0\ &\ 0\\
                    0&x&0&x&0&0&0&0\\
                    0&0&x&x&0&0&0&0\\
                    0&0&0&x&0&0&0&0\\
                    0&0&0&0&x&0&0&x\\
                    0&0&0&0&0&x&0&x\\
                    0&0&0&0&0&0&x&x\\
                    0&0&0&0&0&0&0&x\end{bmatrix}
\end{equation}
where $x$ indicates a possible non-zero position. For each value of $\left(x_9,\ldots,x_{16}\right)$, the decoder scans independently all possible values of $x_4$ and $x_8$, and assigns to them the corresponding 6 ML estimates of the rest of symbols via hard slicers according to equations \eqref{PAM_slicer_1} and \eqref{PAM_slicer_2} at the top of the next page where $r_{i,j}$ denote the entries of the upper-triangular matrix $\Rm$.
\begin{table*}[t!]
\normalsize
\begin{eqnarray}
x^{\text{ML}}_i\vert\left(\hat{x}_4,\hat{x}_9,\ldots,\hat{x}_{16}\right)&=&\text{sign}\left(z_i\right)\times \text{min}\Big[\big\vert 2\ \text{round}\big(\left(z_i-1\right)/2\big)+1\big\vert,\sqrt{M}-1\Big],\ i=1,2,3 \label{PAM_slicer_1}\\ 
x^{\text{ML}}_j\vert\left(\hat{x}_8,\hat{x}_9,\ldots,\hat{x}_{16}\right)&=&\text{sign}\left(z_j\right)\times \text{min}\Big[\big\vert 2\ \text{round}\big(\left(z_j-1\right)/2\big)+1\big\vert,\sqrt{M}-1\Big],\ j=5,6,7\label{PAM_slicer_2}
\end{eqnarray}
where:
\begin{equation*}
z_i=\left(y_i'r_{i,4}\hat{x}_4-\sum^{16}_{k=9}r_{i,k}\hat{x}_k\right)/r_{i,i},\ i=1,2,3,\ z_j=\left(y_j'-r_{j,8}\hat{x}_8-\sum^{16}_{k=9}r_{j,k}\hat{x}_k\right)/r_{j,j},\ j=5,6,7
\end{equation*}
\hrule
\end{table*}
Therefore, instead of evaluating the ML metric $M^8$ times, the sphere decoder needs only $2M^{4.5}$ ML metric evaluations.

A rate-3/2 code that will be denoted $\Xm_{3/2}$ may be easily obtained by puncturing the rate-2 proposed code in \eqref{R2}, and may be expressed as:
\begin{equation}
\begin{split}
\Xm_{3/2}\left(x_1,\ldots,x_{12}\right)=&\Xm_1\left(x_1,\ldots,x_8\right)+\\ &e^{j\phi_{\text{opt}}}\Xm_1\left(x_9,\ldots,x_{12} \right)\Um_{\text{opt}}  
\end{split}
\label{R3/2}
\end{equation}
\begin{table*}
\centering
\topcaption{summary of comparison in terms of worst-case complexity, Min det and PAPR}
\centering
\begin{tabular}{|c|c|c|c|c|c|}
\hline
Code&Worst-case&Min det ($=\sqrt{\delta}$)&\multicolumn{3}{c|}{PAPR (dB)}\\
\cline{4-6} &complexity&for QAM constellations&QPSK&16QAM&64QAM\\
\hline
The proposed rate-1 code&$2\sqrt{M}$&1&0&2.5&3.7\\
\hline
M.Sinnokrot-J.Barry code \cite{LOWPAPR_SSD}&$4\sqrt{M}$&7.11&0&2.5&3.7\\
\hline
Md.Khan-S.Rajan code \cite{CIOD}&$4\sqrt{M}$&12.8&5.8&8.3&9.5\\
\hline
The proposed rate-3/2 code&$2M^{2.5}$&1 (verified for 4-QAM)&3&5.6&6.7\\
\hline
S.Sirianunpiboon \textit{et al.} code \cite{4TX_MPLX_ORTH}&$M^3$&N/A&5.4&8&8.4\\
\hline
P.Srinath-S.Rajan rate-3/2 code \cite{EX_CIOD}&$4M^{2.5}$&12.8&4&6.5&7.7\\  
\hline
The proposed rate-2 code&$2M^{4.5}$&1 (verified for 4-QAM)&2.8&5.3&6.5\\
\hline
P.Srinath-S.Rajan code \cite{EX_CIOD}&$4M^{4.5}$& 12.8 (verified for 4/16-QAM)&2.8&5.3&6.5\\
\hline
Tian Peng Ren \textit{et al.} rate-2 code \cite{R>1}&$M^5$&N/A&4&7&7.66\\
\hline
\end{tabular}
\label{complexity}
\vspace*{0.3cm}
\hrule
\end{table*}
\section{Numerical and simulation results}
In this section, we compare our proposed codes to similar low-complexity STBCs existing in the literature in terms of worst-case decoding complexity, Peak-to-Average Power Ratio (PAPR), average decoding complexity, coding gain, and bit error rate (BER) performance over quasi-static Rayleigh fading channels. One can notice from Table~\ref{complexity} that the worst-case decoding complexity of the proposed rate-1 code is half that of \cite{CIOD} and \cite{LOWPAPR_SSD}. The worst-case decoding complexity of the proposed rate-3/2 code is half that of the punctured rate-3/2 P.Srinath-S.Rajan code \cite{EX_CIOD} and is smaller by a factor of $\sqrt{M}/2$ than the rate-3/2 S.Sirianunpiboon \textit{et al.} code \cite{4TX_MPLX_ORTH}. The worst-case decoding complexity of our rate-2 code is half that of the rate-2 P.Srinath-S.Rajan code \cite{EX_CIOD} and smaller by a factor of $\sqrt{M}/2$ than the rate-2 T.P. Ren code {\it et al.} \cite{R>1}.

Simulations are carried out in quasi-static Rayleigh fading channel in the presence of AWGN for 2 receive antennas. The ML detection is performed via a depth-first tree traversal with infinite initial radius sphere decoder. The radius is updated whenever a leaf node is reached and sibling nodes are visited according to the simplified Schnorr-Euchner enumeration \cite{SE}. 

From Fig.~\ref{2_4_6bpcu_4x2_BER_ML}, one can notice that the proposed rate-1 code loses about 0.6 dB w.r.t to Md.Khan-S.Rajan rate-1 code \cite{CIOD} while offering similar performance to M.Sinnokrot-J.Barry rate-1 code \cite{LOWPAPR_SSD} at $10^{-3}$ BER for several spectral efficiencies namely 2,4, and 6 bpcu. 
\tikzset{every picture/.style={scale=0.65,every picture/.style={}}}
%
%
%
%
\begin{figure}
\begin{tikzpicture}
\begin{semilogyaxis}[%
view={0}{90},
width=4.5in,
height=3.5in,
scale only axis,
xmin=0, xmax=30,
xlabel={SNR per receive antenna},
xmajorgrids,
ymin=5e-07, ymax=1,
yminorticks=true,
ylabel={BER},
ymajorgrids,
yminorgrids,
legend style={at={(0.03,0.03)},anchor=south west,align=left}]

\addplot [
color=blue,
solid,
line width=2.0pt,
mark=triangle,
mark options={solid,,rotate=270}
]
coordinates{
 (0,0.104877609427609)(3,0.0425102693602694)(6,0.0101673400673401)(9,0.00123013468013468)(12,7.45791245791246e-05)(15,2.35690235690236e-06) 
};

\addplot [
color=black,
solid,
line width=2.0pt,
mark=square,
mark options={solid}
]
coordinates{
 (0,0.103625420875421)(3,0.0430978114478114)(6,0.0110079124579125)(9,0.00142845117845118)(12,7.67676767676768e-05)(15,2.02020202020202e-06) 
};

\addplot [
color=red,
solid,
line width=2.0pt,
mark=o,
mark options={solid}
]
coordinates{
 (0,0.0955077441077441)(3,0.0360102693602693)(6,0.00759242424242424)(9,0.000750673400673401)(12,3.26599326599327e-05)(15,8.41750841750842e-07) 
};


\addplot [
color=blue,
solid,
line width=2.0pt,
mark=triangle,
mark options={solid,,rotate=270}
]
coordinates{
 (0,0.238786868686869)(3,0.169682323232323)(6,0.10490404040404)(9,0.0494112794612794)(12,0.0147823232323232)(15,0.00230993265993266)(18,0.000192592592592593)(21,8.08080808080808e-06) 
};

\addplot [
color=black,
solid,
line width=2.0pt,
mark=square,
mark options={solid}
]
coordinates{
 (0,0.234727946127946)(3,0.165047306397306)(6,0.100520707070707)(9,0.0470099326599327)(12,0.0144526936026936)(15,0.00238585858585859)(18,0.000183164983164983)(21,7.07070707070707e-06) 
};

\addplot [
color=red,
solid,
line width=2.0pt,
mark=o,
mark options={solid}
]
coordinates{
 (0,0.228164478114478)(3,0.158097643097643)(6,0.0936114478114478)(9,0.0409478114478115)(12,0.0109351851851852)(15,0.00140892255892256)(18,8.60269360269361e-05)(21,2.69360269360269e-06) 
};


\addplot [
color=blue,
solid,
line width=2.0pt,
mark=triangle,
mark options={solid,,rotate=270}
]
coordinates{
 (0,0.323627104377104)(3,0.264982996632997)(6,0.202095454545454)(9,0.14245404040404)(12,0.0890486531986532)(15,0.0435875420875421)(18,0.0136574074074074)(21,0.00235791245791246)(24,0.000199158249158249)(27,7.91245791245791e-06) 
};
\addlegendentry{\scriptsize Proposed rate-1 code};

\addplot [
color=black,
solid,
line width=2.0pt,
mark=square,
mark options={solid}
]
coordinates{
 (0,0.319842760942761)(3,0.260387037037037)(6,0.196661111111111)(9,0.136701178451178)(12,0.0836452861952862)(15,0.0402631313131313)(18,0.0128710437710438)(21,0.00228838383838384)(24,0.000198989898989899)(27,6.06060606060606e-06) 
};
\addlegendentry{\scriptsize M.Sinnokrot-J.Barry code};

\addplot [
color=red,
solid,
line width=2.0pt,
mark=o,
mark options={solid}
]
coordinates{
 (0,0.316051683501684)(3,0.254868855218855)(6,0.190067676767677)(9,0.130239898989899)(12,0.0780395622895623)(15,0.0352631313131313)(18,0.00982946127946128)(21,0.00137390572390572)(24,9.15824915824916e-05)(27,2.35690235690236e-06) 
};
\addlegendentry{\scriptsize Md.Khan-S.Rajan code};
\end{semilogyaxis}
\draw[line width=1pt,rotate around={-45:(2.8,5.5)}] (2.8,5.5) ellipse (10pt and 20pt); 
\draw (1.5,5.5) node {\footnotesize 2 bpcu};
\draw[line width=1pt,rotate around={-45:(5.4,5.5)}] (5.4,5.5) ellipse (10pt and 20pt); 
\draw (4.2,5.5) node {\footnotesize 4 bpcu};
\draw[line width=1pt,rotate around={-45:(7.8,5.5)}] (7.8,5.5) ellipse (10pt and 20pt); 
\draw (6.7,5.5) node {\footnotesize 6 bpcu};
\end{tikzpicture}%
\caption{BER performance for 4$\times$2 system}
\label{2_4_6bpcu_4x2_BER_ML}
\end{figure}

\begin{figure}
 \subfigure[Average complexity for 4$\times$2 system at 3 bpcu]{
%
%
%
%
\begin{tikzpicture}

\begin{axis}[%
view={0}{90},
width=4.5in,
height=3.5in,
scale only axis,
xmin=0, xmax=16,
xlabel={SNR per receive antenna},
xmajorgrids,
ymin=10, ymax=90,
ylabel={Average number of visited nodes},
ymajorgrids,
legend style={at={(0.97,0.97)},anchor=north east,align=right}]

\addplot [
color=red,
solid,
line width=2.0pt,
mark=o,
mark options={solid}
]
coordinates{
 (0,86.504642)(4,62.522216)(8,35.925268)(12,20.552392)(16,16.625042) 
};
\addlegendentry{\scriptsize P.Srinath-S.Rajan rate-3/2 code};

\addplot [
color=black,
solid,
line width=2.0pt,
mark=square,
mark options={solid}
]
coordinates{
 (0,60.971564)(4,44.871778)(8,27.717194)(12,16.505152)(16,12.842782) 
};
\addlegendentry{\scriptsize S.Sirianunpiboon {\it et al.} code};

\addplot [
color=blue,
solid,
line width=2.0pt,
mark=triangle,
mark options={solid,,rotate=270}
]
coordinates{
 (0,54.632566)(4,41.120254)(8,26.200852)(12,16.550394)(16,13.096362) 
};
\addlegendentry{\scriptsize Proposed rate-3/2 code};

\end{axis}
\end{tikzpicture}%
  \label{3bpcu_4x2_AV_COM_ML}
 }
 \subfigure[Average complexity for 4$\times$2 system at 6 bpcu]{
%
%
%
%
\begin{tikzpicture}

\begin{axis}[%
view={0}{90},
width=4.5in,
height=3.5in,
scale only axis,
xmin=0, xmax=25,
xlabel={SNR per receive antenna},
xmajorgrids,
ymin=0, ymax=800,
ylabel={Average number of visited nodes},
ymajorgrids,
legend style={at={(0.97,0.97)},anchor=north east,align=right}]

\addplot [
color=red,
solid,
line width=2.0pt,
mark=o,
mark options={solid}
]
coordinates{
 (0,631.433512)(3,393.310204)(6,256.213924)(9,176.835864)(12,118.282704)(15,66.953944)(18,34.205776)(21,20.96154)(24,17.181548) 
};
\addlegendentry{\scriptsize P.Srinath-S.Rajan rate-3/2 code};

\addplot [
color=black,
solid,
line width=2.0pt,
mark=square,
mark options={solid}
]
coordinates{
 (0,708.082832)(3,404.782508)(6,238.86846)(9,151.10166)(12,98.121212)(15,57.248264)(18,30.05494)(21,17.717004)(24,13.56094) 
};
\addlegendentry{\scriptsize S.Sirianunpiboon {\it et al.} code};

\addplot [
color=blue,
solid,
line width=2.0pt,
mark=triangle,
mark options={solid,,rotate=270}
]
coordinates{
 (0,461.887476)(3,291.617224)(6,192.653928)(9,135.046588)(12,92.622704)(15,55.812688)(18,30.771756)(21,18.89098)(24,14.332464) 
};
\addlegendentry{\scriptsize Proposed rate-3/2 code};

\end{axis}
\end{tikzpicture}%
  \label{6bpcu_4x2_AV_COM_ML}
 }
 \subfigure[Average complexity for 4$\times$2 system at 9 bpcu]{
%
%
%
%
\begin{tikzpicture}

\begin{axis}[%
view={0}{90},
width=4.5in,
height=3.5in,
scale only axis,
xmin=0, xmax=30,
xlabel={SNR per receive antenna},
xmajorgrids,
ymin=0, ymax=16000,
ylabel={Average number of visited nodes},
ymajorgrids,
legend style={at={(0.97,0.97)},anchor=north east,align=right}]

\addplot [
color=black,
solid,
line width=2.0pt,
mark=square,
mark options={solid}
]
coordinates{
 (0,15345.3112)(3,6364.14904)(6,2469.4372)(9,947.39191)(12,398.53903)(15,205.6286)(18,126.13324)(21,74.07333)(24,37.39408)(27,20.04971)(30,14.25146) 
};
\addlegendentry{\scriptsize S.Sirianunpiboon {\it et al.} code};

\addplot [
color=red,
solid,
line width=2.0pt,
mark=o,
mark options={solid}
]
coordinates{
 (0,6785.28250303031)(3,3136.76715151515)(6,1409.97267272727)(9,665.073012121212)(12,361.359648484849)(15,233.840557575758)(18,158.134048484849)(21,91.4163696969697)(24,44.3488)(27,24.0475939393939)(30,18.0628242424242) 
};
\addlegendentry{\scriptsize P.Srinath-S.Rajan rate-3/2 code};

\addplot [
color=blue,
solid,
line width=2.0pt,
mark=triangle,
mark options={solid,,rotate=270}
]
coordinates{
 (0,5855.9766)(3,2714.10177575758)(6,1221.41805454545)(9,574.058036363636)(12,306.425157575758)(15,193.998581818182)(18,131.476)(21,79.4647575757575)(24,41.7627696969697)(27,23.0742181818182)(30,15.9575696969697) 
};
\addlegendentry{\scriptsize Proposed rate-3/2 code};

\end{axis}
\end{tikzpicture}%
  \label{9bpcu_4x2_AV_COM_ML}
 }
 \caption{Average complexity for 4$\times$2 system at 3, 6 and 9 bpcu}
\end{figure}
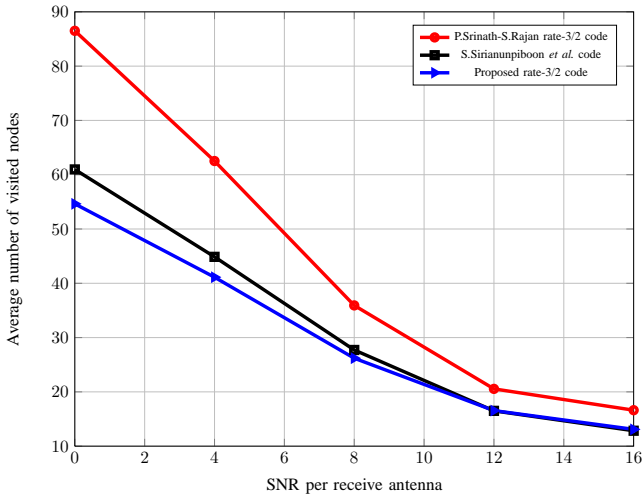
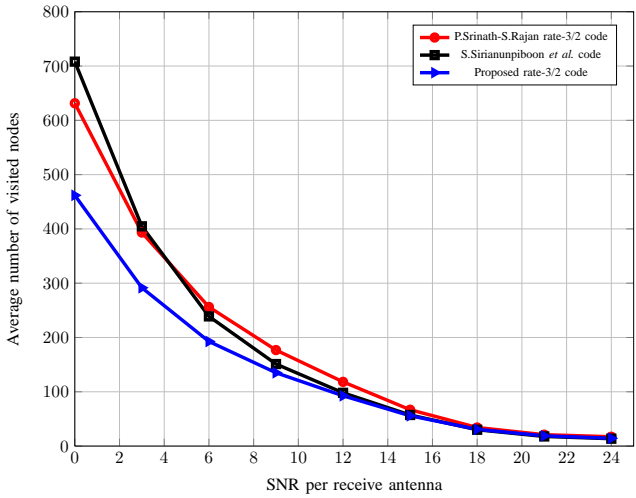
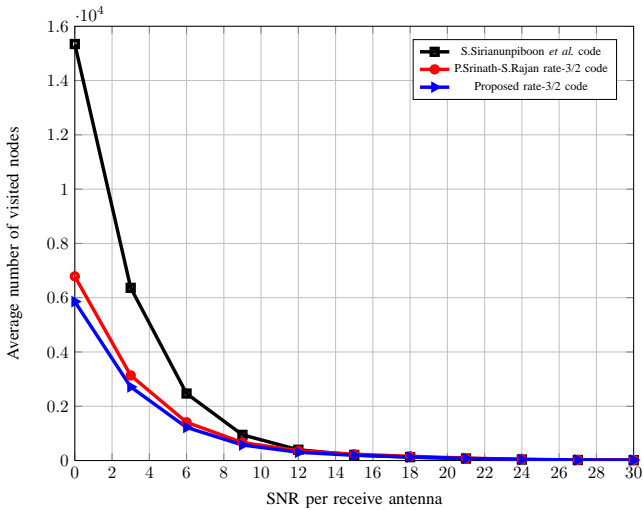

From Fig.~\ref{3_6_9bpcu_4x2_BER_ML}, one can notice that the proposed rate-3/2 code loses about 0.6 dB w.r.t the punctured P.Srinath-S.Rajan code \cite{EX_CIOD} while it gains about 0.4 dB w.r.t S.Sirianunpiboon \textit{et al.} code \cite{4TX_MPLX_ORTH} at $10^{-3}$ BER for several spectral efficiencies namely 3,6, and 9 bpcu. Moreover, from Figs~\ref{3bpcu_4x2_AV_COM_ML}, \ref{6bpcu_4x2_AV_COM_ML}, and \ref{9bpcu_4x2_AV_COM_ML} one can easily verify that the proposed rate-3/2 code maintains its lower average decoding complexity for the considered spectral efficiencies while the average decoding complexity of the S.Sirianunpiboon \textit{et al.} code increases with the size of the underlying constellation.

%
%
%
%
\begin{figure}
\begin{tikzpicture}

\begin{semilogyaxis}[%
view={0}{90},
width=4.5in,
height=3.5in,
scale only axis,
xmin=0, xmax=30,
xlabel={SNR per receive antenna},
xmajorgrids,
ymin=1e-06, ymax=1,
yminorticks=true,
ylabel={BER},
ymajorgrids,
yminorgrids,
legend style={at={(0.03,0.03)},anchor=south west,align=left}]

\addplot [
color=black,
solid,
line width=2.0pt,
mark=square,
mark options={solid}
]
coordinates{
 (0,0.191254666666667)(4,0.0902666666666666)(8,0.0174243333333333)(12,0.000878500000000001)(16,1.1e-05) 
};

\addplot [
color=blue,
solid,
line width=2.0pt,
mark=triangle,
mark options={solid,,rotate=270}
]
coordinates{
 (0,0.174163166666667)(4,0.0751903333333334)(8,0.0130403333333333)(12,0.000637)(16,9.33333333333334e-06) 
};

\addplot [
color=red,
solid,
line width=2.0pt,
mark=o,
mark options={solid}
]
coordinates{
 (0,0.167994666666667)(4,0.068304)(8,0.0103755)(12,0.0003605)(16,2.33333333333333e-06) 
};


\addplot [
color=black,
solid,
line width=2.0pt,
mark=square,
mark options={solid}
]
coordinates{
 (0,0.320659799999999)(3,0.265120933333333)(6,0.2060863)(9,0.144542966666667)(12,0.0774903333333334)(15,0.0235192333333333)(18,0.00320176666666667)(21,0.0001925)(24,6.36666666666666e-06) 
};

\addplot [
color=blue,
solid,
line width=2.0pt,
mark=triangle,
mark options={solid,,rotate=270}
]
coordinates{
 (0,0.3053041)(3,0.2462283)(6,0.1854125)(9,0.122935366666667)(12,0.0600145)(15,0.0163966333333333)(18,0.00212886666666666)(21,0.000134833333333334)(24,5.43333333333333e-06) 
};

\addplot [
color=red,
solid,
line width=2.0pt,
mark=o,
mark options={solid}
]
coordinates{
 (0,0.2991221)(3,0.236412633333334)(6,0.172985133333333)(9,0.109704033333333)(12,0.0493464333333333)(15,0.0117677666666666)(18,0.0012216)(21,5.72000000000003e-05)(24,1e-06) 
};


\addplot [
color=black,
solid,
line width=2.0pt,
mark=square,
mark options={solid}
]
coordinates{
 (0,0.383612222222222)(3,0.3453875)(6,0.298704444444444)(9,0.245746111111111)(12,0.191218611111111)(15,0.138226111111111)(18,0.0810202777777778)(21,0.0285352777777778)(24,0.00453777777777778)(27,0.000316944444444444)(30,1.02777777777778e-05) 
};
\addlegendentry{\scriptsize S.Sirianunpiboon {\it et al.} code};

\addplot [
color=blue,
solid,
line width=2.0pt,
mark=triangle,
mark options={solid,,rotate=270}
]
coordinates{
 (0,0.372853535353535)(3,0.33042962962963)(6,0.279938888888889)(9,0.224825589225589)(12,0.170615993265993)(15,0.117656228956229)(18,0.0626424242424242)(21,0.0195232323232323)(24,0.0029030303030303)(27,0.000190572390572391)(30,5.72390572390572e-06) 
};
\addlegendentry{\scriptsize Proposed rate-3/2 code};

\addplot [
color=red,
solid,
line width=2.0pt,
mark=o,
mark options={solid}
]
coordinates{
 (0,0.368371885521885)(3,0.322677777777778)(6,0.268233164983165)(9,0.209668855218855)(12,0.153187205387205)(15,0.100300673400673)(18,0.0491329966329966)(21,0.0132279461279461)(24,0.00159023569023569)(27,7.82828282828283e-05)(30,2.02020202020202e-06) 
};
\addlegendentry{\scriptsize P.Srinath-S.Rajan rate-3/2 code};

\end{semilogyaxis}
\draw[line width=1pt,rotate around={-45:(3.6,5.5)}] (3.6,5.5) ellipse (10pt and 20pt); 
\draw (2.3,5.5) node {\footnotesize 3 bpcu};
\draw[line width=1pt,rotate around={-45:(6.5,5.5)}] (6.5,5.5) ellipse (10pt and 20pt); 
\draw (5.3,5.5) node {\footnotesize 6 bpcu};
\draw[line width=1pt,rotate around={-45:(9,5.5)}] (9,5.5) ellipse (10pt and 20pt); 
\draw (7.7,5.5) node {\footnotesize 9 bpcu};
\end{tikzpicture}%
\caption{BER performance for 4$\times$2 system}
\label{3_6_9bpcu_4x2_BER_ML}
\end{figure}
 
From Fig.~\ref{4bpcu_4x2_BER_ML}, one can notice that the proposed rate-2 code loses about 0.8 dB w.r.t the P.Srinath-S.Rajan code \cite{EX_CIOD} while gaining about 0.25 dB w.r.t T.P. Ren {\it et al.} code \cite{R>1} at $10^{-3}$ BER. From Fig.~\ref{4bpcu_4x2_AV_COM_ML}, it is easily noticed that our proposed code is decoded with lower average decoding complexity than T.P. Ren {\it et al.} code \cite{R>1} over the entire SNR range while it maintains its lower average decoding complexity w.r.t P.Srinath-S.Rajan code \cite{EX_CIOD} in the low SNR region. 
%
%
%
%
\begin{figure}
\begin{tikzpicture}

\begin{semilogyaxis}[%
view={0}{90},
width=4.5in,
height=3.5in,
scale only axis,
xmin=0, xmax=18,
xlabel={SNR per receive antenna},
xmajorgrids,
ymin=1e-06, ymax=1,
yminorticks=true,
ylabel={BER},
ymajorgrids,
yminorgrids,
legend style={at={(0.03,0.03)},anchor=south west,align=left}]

\addplot [
color=black,
solid,
line width=2.0pt,
mark=square,
mark options={solid}
]
coordinates{
 (0,0.224273666666667)(3,0.158652)(6,0.0837391666666666)(9,0.0254233333333333)(12,0.00377033333333334)(15,0.000285)(18,1.31666666666667e-05) 
};
\addlegendentry{\scriptsize T.P.Ren {\it et al.} rate-2 code};

\addplot [
color=blue,
solid,
line width=2.0pt,
mark=triangle,
mark options={solid,,rotate=270}
]
coordinates{
 (0,0.223282166666667)(3,0.154912166666667)(6,0.0796216666666667)(9,0.0233586666666667)(12,0.003232)(15,0.00022)(18,6.83333333333333e-06) 
};
\addlegendentry{\scriptsize Proposed rate-2 code};

\addplot [
color=red,
solid,
line width=2.0pt,
mark=o,
mark options={solid}
]
coordinates{
 (0,0.212425)(3,0.140056)(6,0.0648498333333334)(9,0.0159573333333333)(12,0.00171816666666667)(15,6.76666666666666e-05)(18,1.16666666666667e-06) 
};
\addlegendentry{\scriptsize P.Srinath-S.Rajan rate-2 code};

\end{semilogyaxis}
\end{tikzpicture}%
\caption{BER performance for 4$\times$2 system}
\label{4bpcu_4x2_BER_ML}
\end{figure}
%
%
%
%
\begin{figure}
\begin{tikzpicture}

\begin{axis}[%
view={0}{90},
width=4.5in,
height=3.5in,
scale only axis,
xmin=0, xmax=18,
xlabel={SNR per receive antenna},
xmajorgrids,
ymin=0, ymax=800,
ylabel={Average number of visited nodes},
ymajorgrids,
legend style={at={(0.03,0.03)},anchor=south west,align=left}]

\addplot [
color=black,
solid,
line width=2.0pt,
mark=square,
mark options={solid}
]
coordinates{
 (0,639.783162666667)(3,507.842653333333)(6,407.760784)(9,327.631370666667)(12,274.674512)(15,247.264885333333)(18,234.305256) 
};
\addlegendentry{\scriptsize T.P.Ren {\it et al.} rate-2 code};

\addplot [
color=red,
solid,
line width=2.0pt,
mark=o,
mark options={solid}
]
coordinates{
 (0,718.100858666667)(3,495.930130666666)(6,311.861562666667)(9,162.163074666667)(12,72.69788)(15,37.296568)(18,27.0822133333333) 
};
\addlegendentry{\scriptsize P.Srinath-S.Rajan rate-2 code};

\addplot [
color=blue,
solid,
line width=2.0pt,
mark=triangle,
mark options={solid,,rotate=270}
]
coordinates{
 (0,552.192589333333)(3,414.106701333333)(6,293.681245333333)(9,183.801045333333)(12,103.808450666667)(15,60.306648)(18,39.7631413333333) 
};
\addlegendentry{\scriptsize Proposed rate-2 code};

\end{axis}
\end{tikzpicture}%
\caption{Average complexity for 4$\times$2 system}
\label{4bpcu_4x2_AV_COM_ML}
\end{figure}

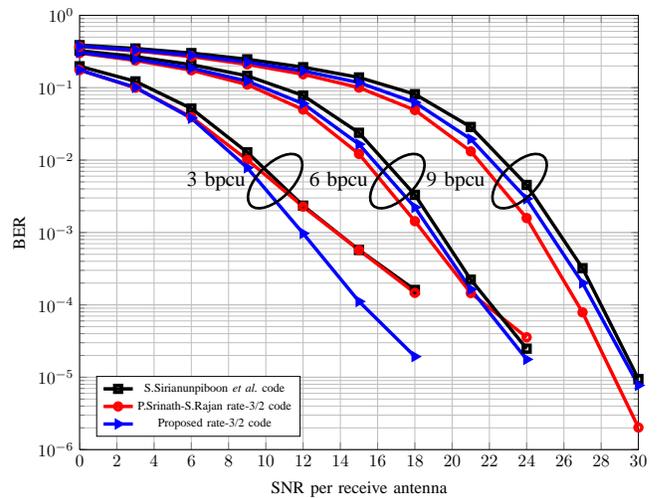
\begin{figure}
\begin{tikzpicture}

\begin{semilogyaxis}[%
view={0}{90},
width=4.5in,
height=3.5in,
scale only axis,
xmin=0, xmax=30,
xlabel={SNR per receive antenna},
xmajorgrids,
ymin=1e-06, ymax=1,
yminorticks=true,
ylabel={BER},
ymajorgrids,
yminorgrids,
legend style={at={(0.03,0.03)},anchor=south west,align=left}]

\addplot [
color=black,
solid,
line width=2.0pt,
mark=square,
mark options={solid}
]
coordinates{
 (0,0.197714333333333)(3,0.122362666666667)(6,0.0516291666666666)(9,0.0128013333333333)(12,0.0023475)(15,0.000573833333333334)(18,0.000160833333333333) 
};

\addplot [
color=red,
solid,
line width=2.0pt,
mark=o,
mark options={solid}
]
coordinates{
 (0,0.175076833333333)(3,0.0999236666666666)(6,0.0393283333333334)(9,0.0100996666666667)(12,0.00227633333333333)(15,0.0005715)(18,0.0001465) 
};

\addplot [
color=blue,
solid,
line width=2.0pt,
mark=triangle,
mark options={solid,,rotate=270}
]
coordinates{
 (0,0.176017166666667)(3,0.1008305)(6,0.0379418333333334)(9,0.007847)(12,0.0009715)(15,0.0001115)(18,1.93333333333333e-05) 
};


\addplot [
color=red,
solid,
line width=2.0pt,
mark=o,
mark options={solid}
]
coordinates{
 (0,0.301201333333333)(3,0.238102)(6,0.174189333333333)(9,0.110413333333333)(12,0.0499006666666667)(15,0.0121778333333333)(18,0.0014325)(21,0.0001455)(24,3.56666666666667e-05) 
};

\addplot [
color=black,
solid,
line width=2.0pt,
mark=square,
mark options={solid}
]
coordinates{
 (0,0.324415166666667)(3,0.268321166666666)(6,0.208161166666667)(9,0.145710333333333)(12,0.0781435)(15,0.0238473333333333)(18,0.00330966666666666)(21,0.000223)(24,2.48333333333333e-05) 
};

\addplot [
color=blue,
solid,
line width=2.0pt,
mark=triangle,
mark options={solid,,rotate=270}
]
coordinates{
 (0,0.306447166666667)(3,0.247044)(6,0.186035666666667)(9,0.123325166666667)(12,0.0603075)(15,0.0165583333333333)(18,0.0022235)(21,0.000164166666666667)(24,1.76666666666667e-05) 
};


\addplot [
color=black,
solid,
line width=2.0pt,
mark=square,
mark options={solid}
]
coordinates{
 (0,0.387095117845118)(3,0.348194612794613)(6,0.300821548821549)(9,0.247017003367003)(12,0.192078451178451)(15,0.138709764309764)(18,0.0814510101010101)(21,0.0286897306397306)(24,0.00453670033670034)(27,0.000321380471380471)(30,9.42760942760943e-06) 
};
\addlegendentry{\scriptsize S.Sirianunpiboon {\it et al.} code};

\addplot [
color=red,
solid,
line width=2.0pt,
mark=o,
mark options={solid}
]
coordinates{
 (0,0.369524410774411)(3,0.323497643097643)(6,0.268675757575758)(9,0.209844781144781)(12,0.15325303030303)(15,0.10031430976431)(18,0.0491279461279461)(21,0.0132119528619529)(24,0.00158215488215488)(27,7.87878787878788e-05)(30,2.02020202020202e-06) 
};
\addlegendentry{\scriptsize P.Srinath-S.Rajan rate-3/2 code};

\addplot [
color=blue,
solid,
line width=2.0pt,
mark=triangle,
mark options={solid,,rotate=270}
]
coordinates{
 (0,0.373684006734007)(3,0.330910942760943)(6,0.280217003367003)(9,0.224947474747475)(12,0.170618181818182)(15,0.117555723905724)(18,0.0625346801346801)(21,0.0195097643097643)(24,0.00292154882154882)(27,0.00020050505050505)(30,7.74410774410775e-06) 
};
\addlegendentry{\scriptsize Proposed rate-3/2 code};

\end{semilogyaxis}
\draw[line width=1pt,rotate around={-45:(4,5.5)}] (4,5.5) ellipse (10pt and 20pt); 
\draw (2.8,5.5) node {\footnotesize 3 bpcu};
\draw[line width=1pt,rotate around={-45:(6.5,5.5)}] (6.5,5.5) ellipse (10pt and 20pt); 
\draw (5.3,5.5) node {\footnotesize 6 bpcu};
\draw[line width=1pt,rotate around={-45:(9,5.5)}] (9,5.5) ellipse (10pt and 20pt); 
\draw (7.7,5.5) node {\footnotesize 9 bpcu};
\end{tikzpicture}%
\caption{BER performance for 4$\times$2 system with timeout sphere decoder}
\label{3_6_9bpcu_4x2_BER_timeout}
\end{figure}

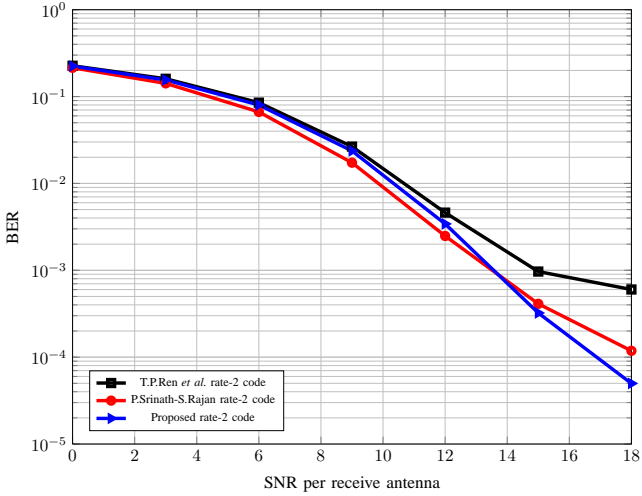
\begin{figure}
\begin{tikzpicture}

\begin{semilogyaxis}[%
view={0}{90},
width=4.5in,
height=3.5in,
scale only axis,
xmin=0, xmax=18,
xlabel={SNR per receive antenna},
xmajorgrids,
ymin=1e-05, ymax=1,
yminorticks=true,
ylabel={BER},
ymajorgrids,
yminorgrids,
legend style={at={(0.03,0.03)},anchor=south west,align=left}]

\addplot [
color=black,
solid,
line width=2.0pt,
mark=square,
mark options={solid}
]
coordinates{
 (0,0.226361333333333)(3,0.160327166666667)(6,0.0850961666666666)(9,0.0265548333333334)(12,0.00461366666666667)(15,0.0009675)(18,0.000601333333333333) 
};
\addlegendentry{\scriptsize T.P.Ren {\it et al.} rate-2 code};

\addplot [
color=red,
solid,
line width=2.0pt,
mark=o,
mark options={solid}
]
coordinates{
 (0,0.214498333333333)(3,0.141883)(6,0.0665693333333333)(9,0.017329)(12,0.002484)(15,0.0004115)(18,0.000118333333333333) 
};
\addlegendentry{\scriptsize P.Srinath-S.Rajan rate-2 code};

\addplot [
color=blue,
solid,
line width=2.0pt,
mark=triangle,
mark options={solid,,rotate=270}
]
coordinates{
 (0,0.223958666666667)(3,0.1553895)(6,0.0800193333333333)(9,0.023706)(12,0.0034235)(15,0.000322833333333333)(18,4.98333333333333e-05) 
};
\addlegendentry{\scriptsize Proposed rate-2 code};

\end{semilogyaxis}
\end{tikzpicture}%
\caption{BER performance for 4$\times$2 system with time out sphere decoder}
\label{4bpcu_4x2_BER_timeout}
\end{figure}

Next, we considered a more practical scenario where a {\it time-out} sphere decoder \cite{TIMEOUT_SD} is employed. In fact, the tree-based search is terminated if a predetermined limit on the number of visited nodes is exceeded and the sphere decoder returns the current codeword estimation. In Fig.~\ref{3_6_9bpcu_4x2_BER_timeout} we fixed a threshold of 50, 500, 5000 nodes count at 3, 6, and 9 spectral efficiencies respectively. It can be verified that our rate-3/2 proposed code outperforms the punctured P.Srinath-S.Rajan code \cite{EX_CIOD} and the S.Sirianunpiboon \textit{et al.} code \cite{4TX_MPLX_ORTH} at high SNR for 3 and 6 bpcu spectral efficiencies. In Fig.~\ref{4bpcu_4x2_BER_timeout} the threshold is fixed at 1000 nodes count. One can easily verify that the proposed code outperforms the P.Srinath-S.Rajan code \cite{EX_CIOD} and the T.P. Ren {\it et al.} code \cite{R>1} code at high SNR. This can be justified by the fact that the maximum number of visited nodes is related to the worst-case decoding complexity which is lower in our proposed codes.
%
%
%
%
%
%
%
%
\section{Conclusions}
In the present paper we have proposed a systematic approach for the construction of rate-1 FGD codes for an arbitrary number of transmit antennas. This approach when applied to the special case of four transmit antennas results in a new rate-1 FGD STBC that has the smallest worst-case decoding complexity among existing comparable low-complexity STBCs. The coding gain of the proposed FGD rate-1 code was then optimized through constellation stretching. Next we managed to increase the rate to 2 by multiplexing two rate-1 codes through a unitary matrix. A compromise between complexity and throughput may be achieved through puncturing the proposed rate-2 code which results in a new low-complexity rate-3/2 code. The worst-case decoding complexity of the proposed codes is lower than their STBC counterparts in the literature. 

According to the simulations results, the proposed rate-1 code loses about 0.6 dB w.r.t to Md.Khan-S.Rajan rate-1 code \cite{CIOD} while offering similar performance to M.Sinnokrot-J.Barry rate-1 code \cite{LOWPAPR_SSD} at $10^{-3}$ BER for several spectral efficiencies namely 2,4, and 6 bpcu. The proposed rate-3/2 code loses about 0.6 dB w.r.t the punctured P.Srinath-S.Rajan code \cite{EX_CIOD} while it gains about 0.4 dB w.r.t S.Sirianunpiboon \textit{et al.} code \cite{4TX_MPLX_ORTH} at $10^{-3}$ BER for several spectral efficiencies namely 3,6, and 9 bpcu. Moreover, the proposed rate-3/2 code maintains its lower average decoding complexity for the considered spectral efficiencies.
 
The proposed rate-2 code loses about 0.8 dB w.r.t the P.Srinath-S.Rajan code \cite{EX_CIOD} while gaining about 0.25 dB w.r.t T.P. Ren {\it et al.} code \cite{R>1} at $10^{-3}$ BER . Our proposed code is decoded with lower average decoding comlexity than T.P. Ren {\it et al.} code \cite{R>1} over the entire SNR range while it maintains its lower average decoding complexity w.r.t P.Srinath-S.Rajan code \cite{EX_CIOD} in the low SNR region. 

Next, we considered a more practical scenario where a {\it time-out} sphere decoder is employed. Our rate-3/2 proposed code outperforms the punctured P.Srinath-S.Rajan code \cite{EX_CIOD} and the S.Sirianunpiboon \textit{et al.} code \cite{4TX_MPLX_ORTH} at high SNR for 3 and 6 bpcu spectral efficiencies.The proposed rate-2 code outperforms the P.Srinath-S.Rajan code \cite{EX_CIOD} and the T.P. Ren {\it et al.} code \cite{R>1} code at high SNR. This can be justified by the fact that the maximum number of visited nodes is related to the worst-case decoding complexity which is lower in our proposed codes
\section*{Appendix A}
\begin{proof}
From the properties of the matrix representations of the Clifford algebra generators over $\mathbb{R}$ \eqref{properties}, it is straightforward to prove that for a matrix $\Am\in\Mc_{2^a}$, if $\Am=\prod^{m}_{i=1}\Rm_{k_i}: 1\leq k_1<\ldots<k_m\leq 2a+1$, then we have:
\begin{equation}
\Am^{H}=(-1)^{m(m+1)/2}\Am
\label{ishermitian}
\end{equation}
moreover, it is easy to prove that:
\begin{eqnarray}
\Rm_l\Am=\left\lbrace \begin{array}{ll} \Am\Rm_l& \begin{array}{ll}m\ \text{odd}&l\in\lbrace k_1,\ldots,k_m\rbrace\\m\ \text{even}&l\notin\lbrace k_1,\ldots,k_m\rbrace\end{array}\\
-\Am\Rm_l&\begin{array}{ll}m\ \text{even}&l\in\lbrace k_1,\ldots,k_m\rbrace\\m\ \text{odd}&l\notin\lbrace k_1,\ldots,k_m\rbrace\end{array}\\\end{array}.\right.
\label{commutativity}
\end{eqnarray}
It is easy to see that $\Gc_1\setminus \Ac \cup \Gc_2\setminus \Bc$ is the set of weight matrices for the Complex Orthogonal Design (COD) for $2^a$ transmit antennas. Thus we need only to prove that the $\Gc_1$, and $\Gc_2$ satisfy \eqref{g-group}. Towards this end let $\Am_k\in\Ac$, $\Bm_l\in\Bc$, $\Cm\in\Gc_1\setminus\Ac$, and $\Dm\in\Gc_2\setminus\Bc$. Let $a=4n$, we have according to Table~\ref{casesB}:
\begin{equation}
\Bm_1=j\prod^{a}_{i=1}\Rm_i
\label{B1}
\end{equation}
and $\Bm_{m/2+1}$ is given by \eqref{Bm} at the top of the next page.
\begin{table*}[t!]
\normalsize
\begin{equation}
\Bm_{m/2+1}=\left\lbrace \begin{array}{ll} j\prod^{a}_{i=1}\Rm_i \prod^{m}_{i=1}\Rm_{k_i}&a+1\leq k_1<\ldots<k_m\leq 2a+1,\ m=4n,\ n'\neq 0\\
\prod^{a}_{i=1}\Rm_i \prod^{m}_{i=1}\Rm_{k_i}&a+1\leq k_1<\ldots< k_m\leq 2a+1,\ m=4n'+2 \end{array}\right.
\label{Bm}
\end{equation}
\setcounter{equation}{37}
\begin{equation}
\Am_m=\left\lbrace \begin{array}{ll} j\prod^{2a+1}_{i=a+1}\Rm_i \prod^{m}_{i=1}\Rm_{k_i}&1\leq k_1<\ldots< k_m\leq a,\ m=4n',\ n'\neq 0\\ 
\prod^{2a+1}_{i=a+1}\Rm_i \prod^{m}_{i=1}\Rm_{k_i}&1\leq k_1<\ldots< k_m\leq a,\ m=4n'+1,\ n'\neq 0\\ 
\prod^{2a+1}_{i=a+1}\Rm_i \prod^{m}_{i=1}\Rm_{k_i}&1\leq k_1<\ldots< k_m\leq a,\ m=4n'+2\\ 
j\prod^{2a+1}_{i=a+1}\Rm_i \prod^{m}_{i=1}\Rm_{k_i}&1\leq k_1<\ldots< k_m\leq a,\ m=4n'+3\\ 
 \end{array}\right.
\label{Am} 
\end{equation}
\hrule
\end{table*}
Consequently, from \eqref{ishermitian}:
\setcounter{equation}{34}
\begin{equation}
\Bm^H=-\Bm,\ \forall \Bm \in \Bc
\end{equation}
and from \eqref{commutativity} one has:
\begin{equation}
\Bm^H\Cm+\Cm^H\Bm=\zerov,\ \forall\Bm \in \Bc.
\end{equation}
On the other hand from Table~\ref{casesA} we have:
\begin{equation}
\Am_1=j\prod^{2a+1}_{i=a+1}\Rm_i
\label{A1}
\end{equation}
and $\Am_m$ is given by \eqref{Am} at the top of the next page. From \eqref{ishermitian}, it follows that:
\setcounter{equation}{38}
\begin{eqnarray}
\Am^H_1&=&\Am_1\\
\Am^H_m&=&\left\lbrace \begin{array}{ll}
\Am_m&m\ \text{even}\\
-\Am_m&m\ \text{odd}
\end{array}\right.
\end{eqnarray}
and from \eqref{commutativity}, we obtain:
\begin{equation}
\Am^H\Dm+\Dm^H\Am=\zerov, \forall \Am\in\Ac.
\end{equation}
Finally, from Eqs~\eqref{commutativity}, \eqref{B1}, \eqref{Bm}, \eqref{A1}, and \eqref{Am}we get:
\begin{equation}
\Am^H\Bm+\Bm^H\Am=\zerov,\ \forall \Am\in\Ac,\Bm\in\Bc.
\end{equation}
It remains to prove that the proposed weight matrices are linearly independent over $\mathbb{R}$.
Towards this end recall from \cite{ANTICOMMUTING} that if $\left\lbrace \Mm_k:k=1,\ldots,2a\right\rbrace$ are pairwise anti-commuting matrices that square to a scalar, then the set:
\begin{equation*}
\begin{split}
\Bc_{2^a}=\left\lbrace\Id\right\rbrace \cup &\left\lbrace \Mm_k:k=1,\ldots,2a\right\rbrace \overset{2a}{\underset{m=2}{\cup}}\\ 
&\left\lbrace \prod^{m}_{i=1}\Mm_{k_i}:1\leq k_1 < \ldots < k_m \leq 2a\right\rbrace
\end{split}
\end{equation*}
forms a basis of $\Mc_{2^a}$ over $\mathbb{C}$. Consequently, the set $\Bc_{2^a} \cup j \Bc_{2^a}$ forms a basis over $\mathbb{R}$. Thanks to the properties of the matrix representations of Clifford algebra generators \eqref{properties}, the set of matrices defined in \eqref{Rbasis} forms a basis of $\Mc_{2^a}$ over $\mathbb{R}$:
\begin{equation}
\begin{split}
\left\lbrace\Id,j\Id\right\rbrace \cup &\left\lbrace \Rm_k,j\Rm_k:k=1,\ldots,2a\right\rbrace \overset{2a}{\underset{m=2}{\cup}}\\
&\left\lbrace \prod^{m}_{i=1}\Rm_{k_i},j\prod^{m}_{i=1}\Rm_{k_i}:1\leq k_1 < \ldots < k_m \leq 2a\right\rbrace
\end{split}
\label{Rbasis}
\end{equation}
But from \eqref{reps}, one has:
\begin{equation}
\prod^{2a+1}_{i=1}\Rm_i=\mp j^{a-1}\Id_{2^a}
\end{equation}
which reduces for the case of $a=4n$ to:
\begin{equation}
\prod^{2a+1}_{i=1}\Rm_i=\pm j\Id_{2^a}.
\label{prod}
\end{equation}
In what follows, we will express the proposed weight matrices in terms of $\Rm_i, i=1,2,\ldots,2a$ thanks to \eqref{prod}. Readily, the set $\Gc_2\setminus\Bc$ becomes $\left\lbrace \Rm_{a+1},\ldots,\Rm_{2a},\pm j\prod^{2a}_{i=1}\Rm_i\right\rbrace$.  After some manipulations, the sets $\Ac$ in Table~\ref{casesA} and $\Bc$ in Table~\ref{casesB} may be re-written as in \eqref{A} and \eqref{B} respectively at  the top of the next page.
\begin{table*}[t!]
\normalsize
\begin{equation}
\Ac=\left\lbrace \pm \prod^{a}_{i=1}\Rm_i \right\rbrace \overset{a-1}{\underset{m=2}{\cup}}\left\lbrace j^{\delta_{\Ac}(a-m)+1}\prod^{m}_{i=1}\Rm_{k_i}:1\leq k_1<\ldots<k_m\leq a \right\rbrace
\label{A}
\end{equation}
\begin{equation}
\begin{split}
\Bc=\left\lbrace j\prod^{a}_{i=1}\Rm_i\right\rbrace &\overset{a-2}{\underset{m=2,4}{\cup}}
\left\lbrace j^{\delta_{\Bc}(m)}\prod^{a}_{i=1}\Rm_i\prod^{m}_{i=1}\Rm_{k_i}:a+1\leq k_1<\ldots<k_m\leq 2a\right\rbrace\\&\overset{a-1}{\underset{m=3,5}{\cup}}
\left\lbrace j^{\delta_\Bc(a+1-m)+1} \prod^{m}_{i=1} \Rm_{k_i}:a+1\leq k_1<\ldots<k_m\leq 2a\right\rbrace 
\end{split}
\label{B}
\end{equation}
\hrule
\end{table*}
Now it can be easily verified that $\Gc_1\cup\Gc_2$ is a subset of the basis in \eqref{Rbasis}. The proofs for other cases of $a$ follow similarly and are therefore omitted.
\end{proof}
\section*{Appendix B}
\begin{proof} The rate of the proposed FGD codes for the case of $2^a$ transmit antennas may be expressed as:
\begin{equation}
R=\frac{2a+2+\vert\Ac\vert+\vert\Bc\vert}{2^{a+1}} 
\end{equation}
However form Table~\ref{casesA}, regardless of $a$ we have:
\begin{equation}
\vert\Ac\vert=1+\sum^{a-2}_{i=1}\binom{a}{i}=\sum^{a-2}_{i=0}\binom{a}{i}=2^a-(a+1)
\end{equation}
On the other hand, from Table~\ref{casesB}, we have for $a$ even:
\begin{equation}
\begin{split}
\vert\Bc\vert&=1+\sum^{a-2}_{i=2,4,\ldots}\binom{a+1}{i}
=1+\sum^{a-2}_{i=2,4\ldots}\binom{a}{i-1}+\binom{a}{i}\\
&=1+\sum^{a-3}_{j=1,3\ldots}\binom{a}{j}+\sum^{a-2}_{i=2,4\ldots}\binom{a}{i}\\
&=1+\sum^{a-2}_{l=1}\binom{a}{l}=\sum^{a-2}_{l=0}\binom{a}{l}=2^a-(a+1)
\end{split}
\end{equation}
where we used the recursion identity:
\begin{equation}
\binom{n}{k}=\binom{n-1}{k-1}+\binom{n-1}{k}
\end{equation}
Similarly, for $a$ odd, we have:
\begin{equation}
\begin{split}
\vert\Bc\vert&=\sum^{a-2}_{i=1,3,\ldots}\binom{a+1}{i}
=\sum^{a-2}_{i=1,3\ldots}\binom{a}{i-1}+\binom{a}{i}\\
&=\sum^{a-3}_{j=0,2\ldots}\binom{a}{j}+\sum^{a-2}_{i=1,3\ldots}\binom{a}{i}=\sum^{a-2}_{l=0}\binom{a}{l}\\
&=\sum^{a-2}_{l=0}\binom{a}{l}=2^a-(a+1).
\end{split}
\end{equation}
Finally, using these relations, we get:
\begin{equation}
R=1.
\end{equation}
Thus concluding the proof.\end{proof}
\section*{Appendix C}
\begin{proof}
The proposed code is 2-group decodable and the corresponding two sub-codes will be denoted by $\Xm_{\rm{I}}=\Xm(x_1,x_2,x_3,x_4,0,0,0,0)$ and $\Xm_{\rm{II}}=\Xm(0,0,0,0,x_5,x_6,x_7,x_8)$ to avoid any ambiguity. The coding gain $\delta_{\Xm}$ is equal to the minimum Coding Gain Distance (CGD) \cite{STC_TH+PR}, or mathematically:
\begin{eqnarray}
\delta_{\Xm} &=&\underset{\sv, \sv' \in \Cc}{\underset{\sv \neq \sv'}{\text{min}}} 
\underbrace{\text{det}\left(\left(\Xm(\sv)-\Xm(\sv')\right)^H\left(\Xm(\sv)-\Xm(\sv')\right)\right)}_{\text{CGD}(\Xm(\sv),\Xm(\sv'))}\nonumber \\
&=&\underset{\Delta\sv \in \Delta \Cc \backslash\lbrace \zerov \rbrace}{\text{min}}
\vert\text{det}\left(\left(\Xm(\Delta\sv) \right)\right)\vert^{2}
\end{eqnarray}
where $\Delta \sv=\sv-\sv'$, $\Delta \Cc$ is the vector space spanned by $\Delta \sv$.\\
Thanks to the quasi-orthogonality structure one has \cite{TCOM12}:
\begin{equation}
\delta_{\Xm}=\text{min}\left\lbrace\delta_{\Xm_{\rm{I}}},\delta_{\Xm_{\rm{II}}}\right\rbrace.
\end{equation}
The coding gain of the first sub-code is expressed as:
\begin{equation}
\delta_{\Xm_{\rm{I}}}=
\left[\left(\Delta x^2_1+\Delta x^2_2+\Delta x^2_3-k^2\Delta x^2_4\right)\left(\frac{2}{1+k^2}\right)\right]^4 
\end{equation}
Choosing $k=\sqrt{\frac{3}{5}}$, the above expression becomes:
\begin{equation}
\delta_{\Xm_{\rm{I}}}=
\left[\frac{\left(5\Delta x^2_1+5\Delta x^2_2+5\Delta x^2_3-3\Delta x^2_4\right)}{5}\left(\frac{2}{1+\frac{3}{5}}\right)\right]^4
\label{det1}
\end{equation} 
where $\Delta x_i=2 n_i,\ n_i \in \mathbb{Z}$. Consider the Diophantine quadratic equation below:
\begin{equation}
5(X^2_1+X^2_2+X^2_3)-3 X^2_4,\ X_1,X_2,X_3,X_4\in \mathbb{Z}
\label{DE1}
\end{equation} 
In order to find a solution we resort to the following theorem \cite{DE}:
\begin{mytheorem}
The equation:
\begin{equation}
f(x)=a_1x_1^2+a_2x_2^2+a_3x_3^2+a_4x_4^2=0.
\end{equation}
is solvable in rational integers iff the coefficients $a_i,i=1,\dots,4$ are such that:\\
if $a_1a_2a_3a_4\equiv 1\ (\text{mod}\ 8)$, then we require $a_1+a_2+a_3+a_4 \equiv 0\ (\text{mod}\ 8)$.
There are no conditions if $a_1a_2a_3a_4\equiv 2,3,5,6,7\ (\text{mod}\ 8)$. In the case $a_1a_2a_3a_4\equiv 4\ (\text{mod}\ 8)$ and $a_1 \equiv a_2 \equiv 0\ (\text{mod}\ 2)$ then if $\frac{1}{4}a_1a_2a_3a_4\equiv 1\ (\text{mod}\ 8)$ it is required that:\\
\begin{equation}
\frac{1}{2}a_1+\frac{1}{2}a_2+a_3+a_4\equiv \frac{1}{2}\left(a_3^2a_4^2-1\right)\ (\text{mod}\ 8).
\end{equation}
No conditions are required if:
\begin{equation}
\frac{1}{4}a_1a_2a_3a_4\equiv 3,5,7\ (\text{mod}\ 8)
\end{equation}
\end{mytheorem}
\noindent Consequently equation \eqref{DE1} equals to 0 iff $X_1=X_2=X_3=X_4=0$, this follows directly from the above theorem as $-3\times 5 \times 5\times 5\equiv 1\ (\text{mod}\ 8)$ with $5+5+5-3=12\equiv \pm 4\ (\text{mod}\ 8)$. Moreover, one has:
\begin{equation}
5(X^2_1+X^2_2+X^2_3)-3X^2_4\neq\pm 1.
\end{equation}
Otherwise, we must have:
\begin{equation}
 3X^2_4\equiv \pm 1\ (\text{mod}\ 5)
\end{equation}
which cannot be true, since the quadratic residues modulo 5 are 0,1 and 4 \cite{TH_NUM}, thus $3X^2_1 \equiv 0,\pm 3\  \text{or}\pm 2\ (\text{mod}\ 5)$. Therefore, we can write:
\begin{equation}
\Big\vert 5(X^2_1+X^2_2+X^2_3)-3 X^2_4\Big\vert \geq 2,\ \forall\left(X_1,X_2,X_3,X_4\right)\neq \zerov 
\end{equation}
The above equality holds for many cases, take for instance $X_1=X_2=1,X_3=X_4=0$.
It is worth noting that the numerator of the expression \eqref{det1} is a special case of the Diophantine equation in \eqref{DE1} as $\Delta x_i=2 n_i,\ n_i \in \mathbb{Z},\ i=1,2,3,4$. Therefore thanks to the above inequality one has:
\begin{equation}
\delta_{\Xm_1}=\left(8/5\right)^4\frac{2^4}{\left(8/5\right)^4}=16. 
\end{equation}
The coding gain of the second sub-code is expressed as:
\begin{equation}
\delta_{\Xm_{\rm{II}}}=
\left[\left(k^2\Delta x^2_5+k^2\Delta x^2_6+k^2\Delta x^2_7-\Delta x^2_8\right)\left(\frac{2}{1+k^2}\right)\right]^4 
\end{equation}
For $k=\sqrt{\frac{3}{5}}$, the above expression becomes:
\begin{equation}
\delta_{\Xm_{\rm{II}}}=
\left[\frac{\left(3\Delta x^2_5+3\Delta x^2_6+3\Delta x^2_7-5\Delta x^2_8\right)}{5}\left(\frac{2}{1+\frac{3}{5}}\right)\right]^4.
\label{det2}
\end{equation} 
Consider the Diophantine quadratic equation below:
\begin{equation}
3(X^2_5+X^2_6+X^2_7)-5X^2_8,\ X_5,X_6,X_7,X_8\in\mathbb{Z}
\label{DE2}
\end{equation} 
It is easy to verify from {\bf Theorem 1.} that the above equation equals 0 iff $X_5=X_6=X_7=X_8=0$ as
$-3\times 3 \times 3\times 5\equiv 1\ (\text{mod}\ 8)$ with $3+3+3-5=4\equiv  \pm 4\ (\text{mod}\ 8)$.\\
However, we have: 
\begin{equation}
\Big \vert 3(X^2_5+X^2_6+X^2_7)-5X^2_8\Big \vert \geq 1,\ \forall(X_5,X_6,X_7,X_8)\neq \zerov 
\end{equation}
The  above inequality holds for instance by taking $X_5=0,\ X_6=X_7=X_8=1$. 
By noting that the nominator in expression \eqref{det2} is a special case of the Diophantine equation \eqref{DE2} as $\Delta x_i=2 n_i,\ n_i \in \mathbb{Z},\ i=5,6,7,8$, then thanks to the above inequality we have:
\begin{equation}
\delta_{\Xm_2}=\left(4/5\right)^4\frac{2^4}{\left(8/5\right)^4}=1
\end{equation}
and thus $\delta_{\Xm}=1$.
\end{proof}

\bibliography{STBCs}
\bibliographystyle{ieeetr}
\end{document}